**The Moon Illusion explained by the Projective Consciousness Model**
Running head: The Moon Illusion & Projective Consciousness

David Rudrauf[1*], Daniel Bennequin[2*] & Kenneth Williford[3*]

1. FAPSE, Section of Psychology, Swiss Center for Affective Sciences, Campus Biotech, University of Geneva, Geneva, Switzerland
2. Department of Mathematics, IMJ, University of Paris 7, Paris, France
3. Department of Philosophy and Humanities, University of Texas, Arlington, USA

* The authors equally contributed to this report.

Corresponding author:

Prof. David Rudrauf
Campus Biotech
Chemin des Mines, 9
1202 Genève, Suisse
MMEF Lab (Director)
Tél: +41 22 379 09 31
Fax: +41 22 379 06 10

**The Moon often appears larger near the perceptual horizon and smaller high in the sky though the visual angle subtended is invariant. We show how this illusion results from the optimization of a projective geometrical frame for conscious perception through free energy minimization, as articulated in the Projective Consciousness Model. The model accounts for all documented modulations of the illusion without anomalies (e.g., the "size-distance paradox"), surpasses other theories in explanatory power, makes sense of inter- and intra-subjective variability vis-à-vis the illusion, and yields new quantitative and qualitative predictions.**

**Introduction**

The structure of consciousness remains one of the greatest scientific puzzles. Here we support the hypothesis that consciousness is framed by projective geometry and dynamically calibrated to resolve sensory uncertainty by accounting for one of the oldest documented unexplained perceptual illusions, the Moon Illusion. Our account is based on an independently derived mathematical model, the Projective Consciousness Model (PCM)(*1*). (Supplementary Information §1).

Euclidean 3-space frames sensorimotor integration, but the PCM entails that conscious perception, imagination, and action control are framed by projective geometry, which extends this 3-space and spaces for action (*2*) with a plane at infinity and the sets of standard isometries with point transformations preserving relations of incidence of lines and planes. Projective frames are capable of integrating points of view dynamically and are calibrated for subjective rendering in the PCM following principles of Bayesian inference and free energy (FE) minimization (*1*, *3*), based on prior beliefs and sensory evidence.

We show that when observing the Moon (especially when full and bright), the most probable projective frame, that which maximizes the use of information, is heavily dependent on the Moon's perceived elevation and the availability of sensory information. We demonstrate mathematically and verify with simulations that on these assumptions the apparent diameter of the Moon tends to be larger when near the horizon (with "landscape" information available) and smaller when higher, in a manner that accounts for reported individual and environmental variability (*4-9*).

The phenomenon, which also affects the perception of other celestial objects, has posed a puzzle since antiquity (*9*) and is considered an illusion because the visual angle A subtended by the Moon remains the same irrespective of its position in the sky. The illusion depends on the salience of depth cues (*5*, *9*), is diminished when the Moon is seen from upside-down through one's legs (*10*), and is maintained in pictures (*11*).

Taking Account of Distance (TAD) theories (*4-6*, *12-14*) hypothesize that perceptual systems assume the diameter of the Moon to be proportional to its estimated distance in accordance with trigonometry (the "size-distance invariance hypothesis" or SDIH). When assumed to be farther away, the Moon would thus be expected to display a larger diameter. Yet, more subjects judge the Moon larger and *closer* when near the horizon and smaller and *farther away* when elevated (*12*), a situation dubbed the "size-distance paradox". TAD theorists have invoked introspective error to resolve the paradox (*12*, *15*), sometimes bolstered by considerations about visual processing streams (*16*). But why should only introspected diameter correctly match unconsciously estimated quantities, and why would only some subjects suffer from this dissociation?

By contrast, the PCM offers a generative model of the illusion at the conscious perceptual level, accommodates the contextual factors implicated in its occurrence, is immune to the objections to TAD theories, and surpasses previous accounts based on the geometry of visual space (*17, 9*).

**Preliminaries**

Perceptually estimating relative sizes and distances is adaptively crucial; there are conscious perceptions of the relative sizes and distances of objects, though no direct perceptions of metrical values. This is consistent with a projective geometrical model in which there is no canonical notion of distance or ratios of distances, but only relations of incidence, preserving alignments and crossings. Nevertheless, some metrics are more natural than others in this geometry because they have the largest number of symmetry dimensions (n = 6): a Euclidean metric in the restricted affine model ($E_3$) and a spherical metric (Fubini-Study) for the full projective space ($PSO_4$). (For technical details see Supplementary Information §2.)

Several studies (e.g., *18*) have established that all sufficiently far objects are judged as "at infinity" when no indication of relative distance is available. The Moon (constellations, etc.) can be assumed to be "at infinity" in a projective frame, which does not preclude the object's appearing nearer to or farther from the observer.

Physically, the Moon M, seen from a point O, subtends a solid angle A. If M is moved to M' maintaining Euclidean distance from O (e.g., from an elevated position to near the horizon), the visual angle A' has the same aperture. The identity of the subtended visual angles does not by itself allow one to infer that apparent distances d and d' or diameters D and D' are identical.

We assume that there exists a projective frame F that maximizes the use of available sensory information and minimizes the divergence from priors (beliefs and preferences) through the minimization of variational free energy.

Relevant variables and parameters (see Supplementary Information §2) are: the visual angle (HM) between the horizon H and the Moon M, the visual angle A subtended by the object M, depth cues, characteristics of light, landscapes and objects. Prior parameters include the probabilities of environmental features and a default projective frame $F_0$. The states of the variables are modified by projective transformations T, integrating sensory evidence and an unconscious world model R in memory into the Field of Consciousness (FoC). (See Supplementary Information §1.) Judgements on the apparent distance d and diameter D of M issue from this process.

Our explanation of the Moon Illusion consists in proving that the projective frames $F=T(F_0)$ and $F'=T'(F_0)$, which minimize free energy in their situations, entail that the apparent diameter D of an elevated Moon is smaller than the apparent diameter D' of a Moon near the horizon. From this, most observers will infer that distance d is larger than distance d' and experience the Moon as both larger and closer under F'. However, it is possible on this model for observers to infer from different priors that the Moon is both larger and farther away (*9,12-13*). Figure 1 illustrates the overall projective setup and argument.

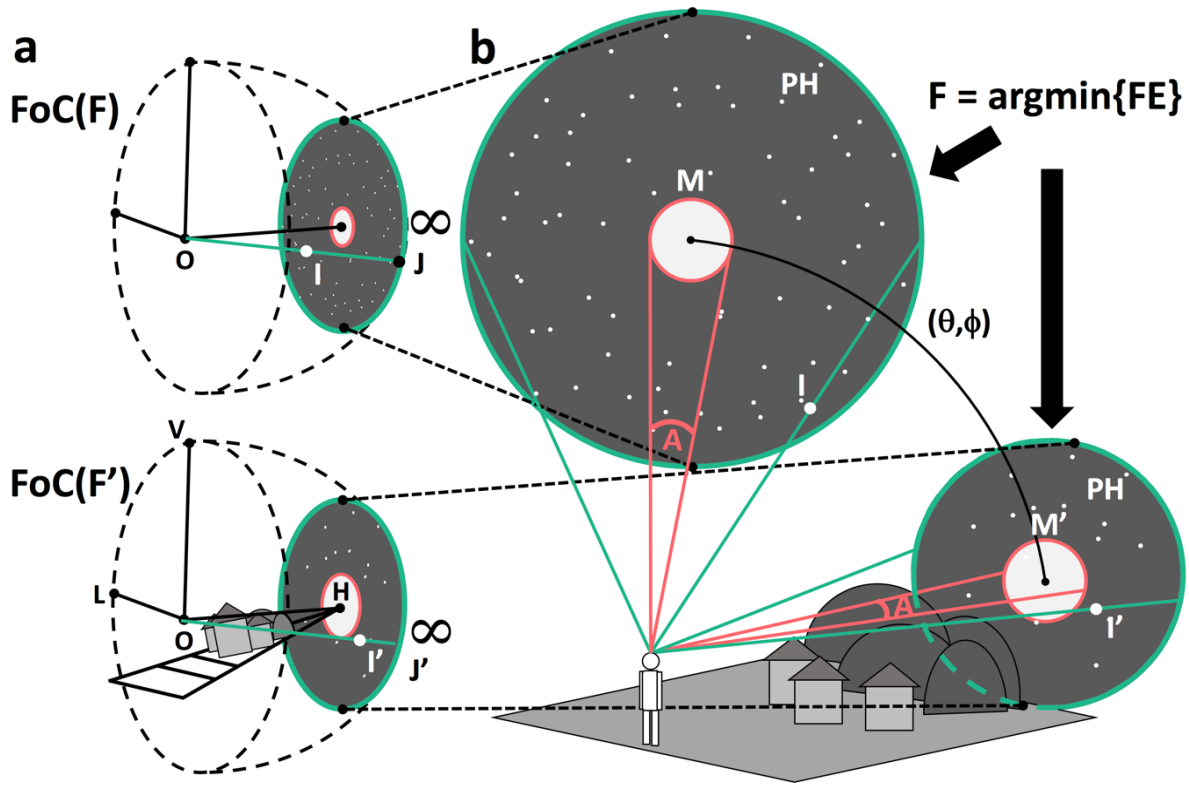

**Fig. 1. Projective setup and argument**
**a.** Fields of Consciousness (FoC) for the Moon high in the sky (top) and on the horizon (bottom) from respective projective frames F and F'. The Moon is at projective infinity. (See text.) The Moon appears smaller high in the sky, larger near the horizon. **b.** Conditions of observation and influences on the frames and their metrics. An observer (manikin) looks at the Moon M high in the sky and M' near the horizon at elevation θ and azimuth ϕ. The visual angle A remains the same. Free energy minimization yields the frame F (or F') with projective transformations for the plane at infinity T (or T'), making maximal use of information. The point I in projective frame F implies a broader projective scope, and I' in frame F', a smaller one, which calibrates the internal metrics of the invariant projective plane at infinity defining the FoC and induces the illusion.

**Default projective frame**

The main ingredient in our explanation is the selection of the projective frame $F=T(F_0)$. Any projective frame for a 3-space must contain five points such that no four of them belong to a plane. (See Supplementary Information §2.)

The standard default frame $F_0$ (itself unconscious but subtending the FoC (*1*)) is defined by the center O (the "point of view"), by three points at infinity, namely, H on the horizon in front, L (or R) laterally, V vertically, and by a fifth point, I, in the finite ambient affine 3-space. The point I behaves as a control point modulating the spatial scope of information integration. H, L, V and the point J projected to infinity along OI define a frame on the plane at infinity $P_\infty$. Standard metrics g (e.g., Euclidean or elliptical) can be adapted to $F_0$.

On the PCM, the perceptual processing of an object like the Moon is as follows (Fig. 1): 1) Constrained by priors, the brain tries to minimize FE by maximizing the use of available information about the surroundings (quantified by an entropy), yielding a projective frame that is optimal for the discrimination of objects and integration of panoramic vision (see formula Supplementary Information §2). 2) The frame selection modifies the standard coordinates, altering apparent relative distances in $P_\infty$. 3) The metric shapes conscious perception, including the apparent size D of the Moon. 4) Priors can be further used, automatically or deliberatively, to attempt to resolve remaining perceptual or conceptual uncertainty (e.g., about the Moon's apparent distance).

**High Moon**

When looking at the Moon M high in the sky, environmental perspective cues (e.g., about relative distances) are generally absent in the observer's visual field, save perhaps the prior that M is beyond the maximum distance of binocular and accommodation discrimination (see *18*). FE minimization leads to an enlargement of the projective scope, since, in such an impoverished context, a larger area offers a greater potential for sampling information (see Supplementary Information §2). This induces a transformation $F = T(F_0)$, moving $I=T(I_0)$ more laterally than $I_0$ in $F_0$ (Fig. 1). Moreover, as the Moon moves towards the zenith, the head of the observer has to be inclined upward; thus $V=T(V_0)$ is shifted in the direction opposite H with respect to $V_0$ in $F_0$. This affects the metric of the FoC, reducing the apparent size of the Moon compared to $F_0$. *Proof*: The angular metric at infinity in the frame F of the FoC is isomorphic to the angular metric in the default frame $F_0$; for instance, in the FoC, $HJ_0$ and $HV_0$ are processed as equal to HJ and HV. But since the real view angle, A, of M remains invariant, it represents a smaller arc in F than in $F_0$. The Moon is thus perceived as smaller. Note that the presence of buildings or other elevated cues is known to mitigate this effect of apparent size reduction (*9*). This can be explained by a smaller displacement of the point I (see Supplementary Information §2).

**Horizon Moon**

When looking at a Moon M' appearing low on the horizon, above the line HL, geometrical priors and available perspective "landscape" cues constrain the choice of the projective transformation T', as more information is present in a narrower region of space. Free energy minimization yields a projective frame F' with a focus on a narrower solid angle around the line OH so that the point I' (and its projection J' at infinity) are closer to the Moon. The arc HV in the FoC has not changed, but the angle HJ' is now slightly smaller than in the default frame, which enlarges the apparent relative size D' of the Moon. *Proof*: The projection of the arc HJ' in the FoC is isomorphic to the arc $HJ'_0$, but the visual angle A is unchanged (A'=A). Therefore, it now constitutes a larger part of the angle HJ, and the Moon is accordingly perceived as larger.

The closer the Moon is to the horizon, the closer the point J' is to H, and the bigger the Moon's apparent size D'. The mathematical derivation (Supplementary Information §2 and Fig. 2) shows a non-linear dependency of D' on the angle HJ'. Note that when comparing arcs at infinity, a prior preference for transformations preserving angles can be assumed. The deformed metric is likely to give the Moon's disc a circular shape consistent with visual angles. The differential displacement of I in the direction of H and L is able to preserve the circular shape. However, an elliptical shape is possible in the general model and might relate to documented effects on halos around the Moon (*9*).

**Active vision and binocular disparity**

Eye convergence, binocular disparity and depth accommodation have the effect of considerably augmenting available information from lateral cues at far distances in the direction of the Moon. On the PCM, this shifts I' toward the line HL and thus augments the illusion. Conversely, monocular vision, eliminating disparity (though not accommodation), which is expected to have the opposite effect on I', is known to strongly attenuate the illusion (*9*).

**The Moon in pictures**

The PCM can also explain why and how the Moon Illusion occurs when making perceptual inferences from planar pictures. The projective frame has to be adapted to the figure's perspective cues, as if a human observer were contained in the world of the picture, which is interpreted as 3-dimensional (not 2D) (*19*). Substitution of points of view and rescaling of frames are a core feature of projective geometry. Following the same information theoretic arguments outlined above, the choice of a projective frame can induce the illusion as a result of the choice of the equivalent of point I. Note that the use of such a frame inside pictures is compatible with the use of another frame for the observer looking at the picture, the different sub-spaces being distinguished in the FoC.

**Seeing the Moon upside down through the legs**

When the Moon is low on the horizon and seen by subjects bending over to look through their legs (*10*), H, M' L, and V can be all assumed to be nearly coplanar. Such a projective setup induces a higher degeneracy of the metric in the plane at infinity, thus flattening the Moon and reducing the Moon Illusion, as empirically documented (*10*). As shown in (*10*), inversion of the visual scene, even in pictures, reduces the illusion. This can be explained by the difficulty of transforming the metric when OV is reversed.

**Results**

**Simulations**

We applied the above principles and quantitative developments (see Methods and Supplementary Information §2 & (*20*)) in simulations demonstrating the illusion and its relation to key model parameters (Fig. 2).

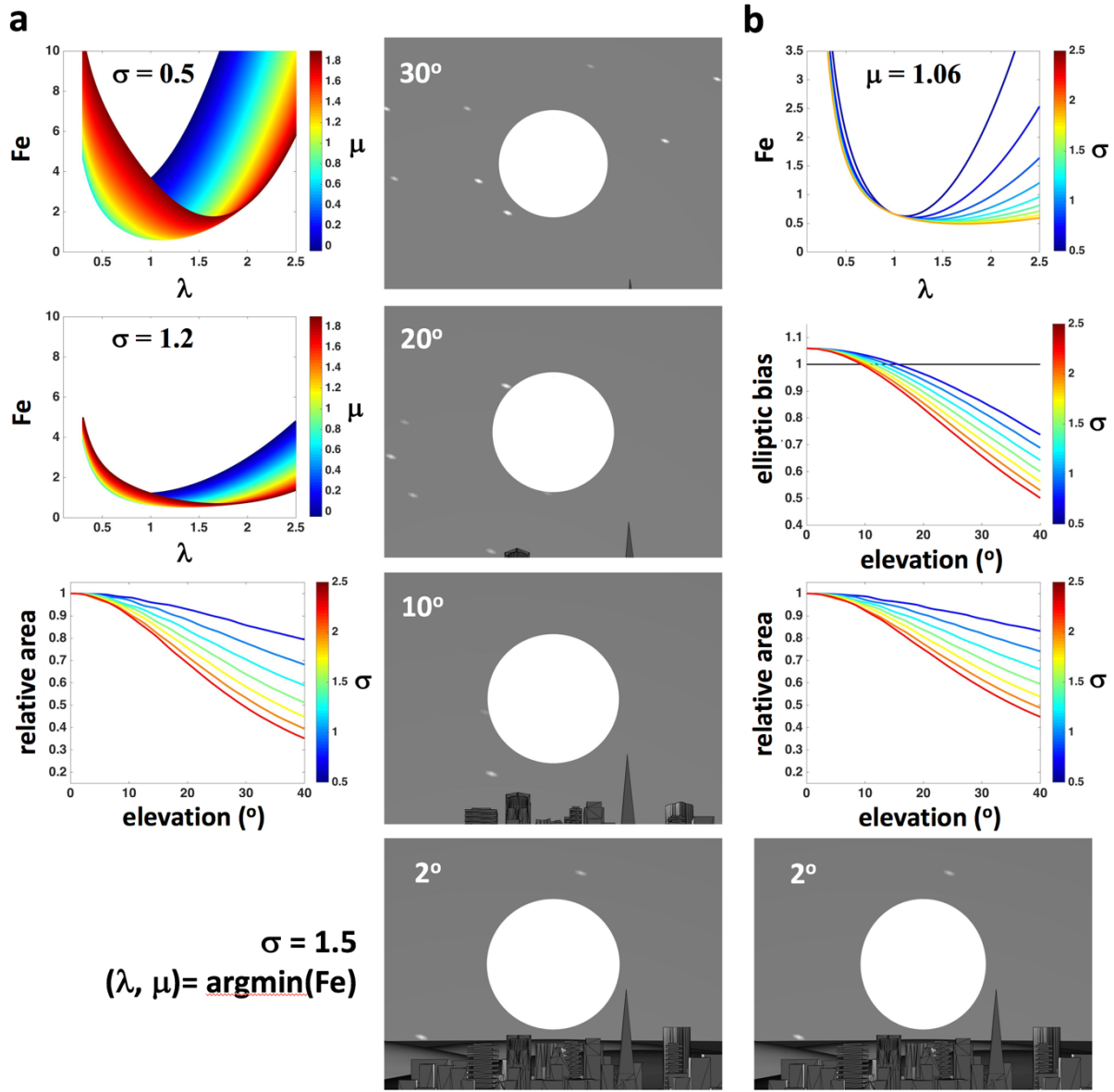

**Fig. 2. Generative projective model of the Moon Illusion.**
**a.** *Left-Tier*: Conformal constraint. Charts representing relations between parameters in the model. *Top* and *Middle*: FE as a function of $\lambda$, $\mu$ and $\sigma$, featuring a strictly convex function guaranteeing a unique solution. *Bottom*: relative area (normalized by the area at 0º elevation) of the perceived Moon as a function of elevation (in degrees) and $\sigma$, demonstrating a range of possible magnification ratios. *Right-Tier*: 2-dimensional planar rendering of the projection of a world model (including the Moon at projective infinity) in projective space, as a function of elevation (in degrees), given parameters: $\sigma$ (influences control point I (see text)), $(\lambda, \mu) =$ *argmin*(Fe). **b.** No conformal constraint. *Top*: FE as a function of $\lambda$ and $\sigma$, with fixed $\mu =$ 1.06. *Middle first*: elliptic bias (ratio of vertical over horizontal diameter) as a function elevation and $\sigma$. *Middle second*: apparent relative area of the perceived Moon as a function of elevation (in degrees) and $\sigma$. *Bottom*: 2-dimensional planar rendering at elevation 2º (See Supplementary Information §2 & (*20*)).

**Further evidence from Virtual Reality**

In order to acquire additional evidence bearing on the model, we performed an initial experiment in Virtual Reality over a small sample (N=6), focusing on specific quantitative predictions (*20*) (Fig. 3) (see Methods). The results demonstrated a relation between elevation and the perceived relative area of the moons (as compared to a reference moon at 2º), consistent with the Moon Illusion, depending on the presence of environmental information. On average, the results fit the nonlinear functions predicted by the PCM better than a linear model (LM) (PCM: $t(6) = 26.3$, $p = 2 *10^{-7}$ versus LM: $t(6) = 1.15$, $p = 0.0004$; relative $R^2 = 0.86$; AIC = - 14.73 versus AIC = -2.76; BIC = -3.98 versus BIC = 0.01). Though not systematic in occurrence and direction and quite variable, we also found some evidence of elliptical deformations of the moons' shapes (Fig. 3b), which were on average significantly larger when environmental information was present than when it was not ($t(4) = 3.63$; $p = 0.02$). Furthermore, 4 of the 5 participants completing the task of elliptical assessment, reported, though irregularly, elliptical deformations of "secondary moons" presented laterally (cf. 3), which is a unique signature of the model when a global conformal prior is not dominating.

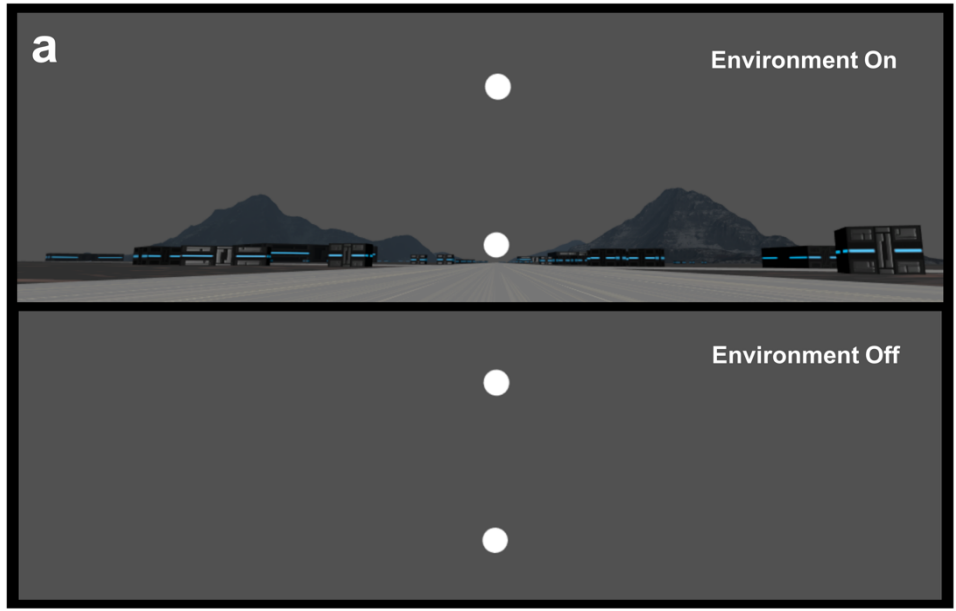

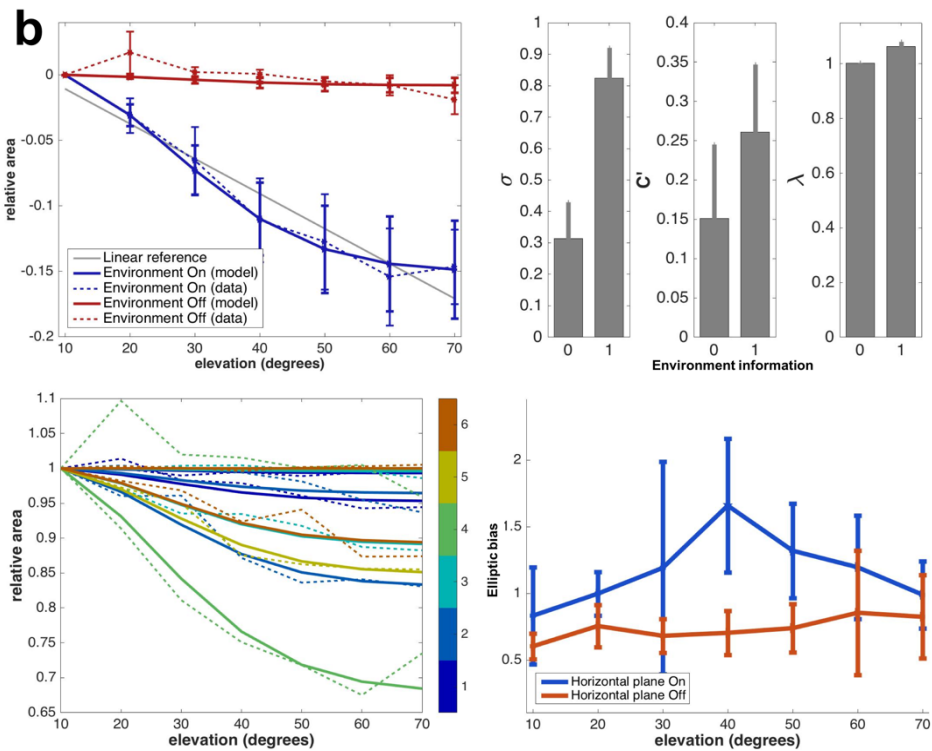

**Fig. 3. Additional empirical evidence from Virtual Reality.**
For methods see (*20*). **a.** Virtual Reality (VR) scenes for conditions: environment On versus Off, displaying a reference moon (near the horizon) and a target moon (at 20° elevation). The participants' task was to change, if warranted, the perceived size of the reference moon to make it match that of the target moons at various elevations. **b.** Result charts. (Error bars are standard errors). *Top-Left*: between-participant average relative perceived area as a function of elevation and condition. With the environment On (blue) (versus Off (red)), on average the empirical perceived areas (dashed curves) decreased (versus did not decrease) with elevation, indicating an effective Moon Illusion in VR, depending on environmental cues. On average, the PCM-predicted, nonlinear curves (continuous lines) demonstrated a good fit with the data, above that of a linear model (grey line). *Bottom-Left*: fitting of individual (participant 1 to 6 (see color bar)) empirical data (dashed curve) and PCM curves (continuous lines). *Top-*

*Right*: average PCM parameters, σ, C', and λ, estimated from empirical data, as a function of the presence [1] or absence [0] of environmental information; corresponding to estimates of the calibration of the participants' FoC frames. *Bottom-Right*: between-participant average elliptic bias as a function of elevation and condition.

**Discussion**

On the PCM, the Moon Illusion is a perceptual phenomenon that results from a projective form of Bayesian inference that frames consciousness. It is induced by the process of the subjective rendering of the FoC, which starts with the maximization of the use of information (through FE minimization) and results in the calibration and application of a projective transformation T to an unconscious world model R. Metrics can be adapted to the finite and infinite compartments in a seamless way, yielding a unique coherent first-person perspectival experience as solution (Supplementary Information §1). At infinity, the projective transformation assumed for the Moon deforms the plane at infinity proportionally to the expected information, which is achieved without affecting the real Moon's visual angle, thus contracting the perceived Moon's relative size when high in the sky and dilating it when low (Supplementary Information §2). Our approach makes quantitative predictions based on explicit parameters about the enlargement of the Moon near the horizon, for instance distinguishing "normal" illusions (a factor > 1 to < 2) from "super" illusions (factor ≥ 2) (*4-8*).

Putative explanations of the Moon Illusion fall into three main types: those appealing to (i) "external physical reasons" (atmospheric refraction, magnification, change in actual distance); (ii) entoptic, optical, and oculomotor processes (faulty accommodation, pupil size under low illumination, binocular disparity); and (iii) "perceptual size changes owing to scaling mechanisms within the brain" (see *9*).

Theories of the first two types can be abandoned (see *9* for comprehensive treatment). The third type, which includes our proposal, TAD theories, and other geometry-based models, still stands. A number of contextual factors impact the Moon Illusion: relative sizes of objects, terrain effects, vergence commands, visual angle, posture, aerial perspective and color (see *9*). The PCM can accommodate the role of basic visual parameters, such as binocular disparity (see *21-23*), as well as the integration of contextual factors. Our model eliminates the "size-distance paradox" plaguing TAD theories, since it implies that D is first perceived and that d is only secondarily inferred, based on priors (e.g., that a constant object should appear bigger when closer) or on lateral cues or accommodation. More generally, it can account for the large inter- and intra-subjective variability affecting apparent D and d.

Heelan (*17*) proposed a finite hyperbolic geometrical model of visual space (i.e., with a negative constant curvature). Though mathematically deep, the model is *ad hoc* if not circular (see (*9*)) as it arbitrarily selects a geometry and a set of fixed parameters to reproduce the illusion. By contrast, the PCM posits that the geometry underlying spatial consciousness is larger than any metrical geometry and tends towards a projective extension of affine geometry, including a notion of variations of points of view, affording great flexibility and complexity. It does not assume fixed structures for measurement but makes metrical properties dependent on active Bayesian inference, motivated by the optimal integration of priors and sensory evidence and accounting for the way in which the world is sampled. Measurements in different planes are independent; sizes and distances are computed independently of each other. Nevertheless, projective geometry can naturally accommodate spherical or Euclidean metrics (at finite distances and infinity), which are perceptually plausible. It can also incorporate hyperbolic metrics, though in the present context this would be with little ecological validity, as these can make parallel lines hyper-parallel and exponentially divergent toward infinity. We have developed a second version of the general model (Supplementary Information §2), where the Moon is assumed to belong to a sphere at a finite distance and where induced transformations are restricted to preserve angles. This second version of the model remarkably coincides with Heelan's model based on the ideal sphere of hyperbolic space. As in Heelan's, the shape of the Moon is always round (conformal). Then the PCM can be compatible with conformal constraints, but it also offers possible non-conformal solutions, inducing elliptical deformations. Our initial results in Virtual Reality suggest the presence of some deviations from conformality, which, if confirmed, would in and of themselves exclude Heelan's model. But more fundamentally, the PCM has much broader explanatory and predictive power (*1*). For instance, in this context, it also provides an explanation of the flattened dome "sky illusion", closely associated with the Moon Illusion (*9*): When looking up at the sky, the point I is displaced laterally, and thus the covered region around the zenith corresponds to a smaller region in the standard default frame. Hence, the distribution of curvature is not perceptually uniform (it is smaller at the zenith and larger at the horizon). More generally, the PCM explains other types of perceptual illusions and provides a psychological inspired generative model of active inference, which encompasses and unifies the frames for perception, imagination and motor programming, embedding them in a general algorithm of global optimization of multimodal information (*1*). If we abandon SDIH and Heelan's model, we need not thereby "abandon geometry" (*9*).

In (*1*) we discussed the possible functional neuroanatomy of the PCM. Here, we further hypothesize that grid cell adaptation (*24-25*) could be linked to transformations of the projective frame, in particular to the extension of grid fields due to the displacement of the point I in the frame. Spatial representations and navigation are supported by place, head direction, grid and boundary cells, which are governed by spatial frames dependent on external cues, in particular "at infinity" (*24-25*), and found in the para-hippocampal region in rodents and also in humans, especially for the visual field (*26*), featuring a larger context-dependent flexibility (*27*).

As Westheimer (*28*) states, the Moon Illusion results from many hidden neuronal activities supporting perception and action and thus involves much more than a geometry, even a projective one, in its overall generative mechanism. However, shifting to a projective space for generating consciousness and, adapting a variable projective frame dynamically through FE minimization, constitute crucial steps in the explanation of the illusion. In turn, the explanation of the Moon Illusion by the PCM provides further support for the validity of the model.

**Methods**

The PCM model principles were applied to obtain a generative model of the Moon Illusion, which was used for both simulations and analyses of empirical results (see Supplementary Information §2 for detailed mathematical definitions and derivations).

**Simulations**

The simulations presented in Fig. 2 were implemented and run using Matlab (MathWorksTM), applying the relevant formula introduced in Supplementary Information §2.
The world model included a series of triangulated meshes: a sphere of radius $OH = 1$ centered at 0 used as a projective plane at infinity; a horizontal ground plane XY; "mountains" based on Matlab's "peaks" function; a model of a city from an online freely available 3-dimensional model; and a small sphere (diameter 0.03) representing the Moon and projected on the sphere at infinity in the direction OH. The coordinate system was: OH = x, OL = y, OV = z.

The parameters of the simulation were varied across the following ranges: elevation of the Moon = [0 2 5 10 15 20 25 30 35 40] degrees; $\sigma$ = [0.5: 0.22 :2.5]; $\lambda$ = [0.3 : 0.0046 : 3.5]; $\mu$ = [0.3 : 0.0046 : 3.5]. We used C = 0 for non-conformal solutions, and C = 6 for conformal solutions. We used C' = 1, and $\nu$ = 1. We used for $\Lambda$ = log($\alpha\gamma$) + log($\omega\kappa$): $\alpha$ = 1, $\gamma$ = 2, $\omega$ = 2, $\kappa$ = 2.

The default frame $F_0$ was derived from V(4) in standard coordinates: [1 0 0 0; 0 1 0 0; 0 0 1 0; 0 0 0 1], with I = c*(V4(1,:)+V4(3,:)-V4(2,:)) (with c = 1), defining P = I*V(4)$^{-1}$, so that $F_0$ was equal to V(4) row multiplied by P elementwise. A transformation $T_{persp}$ was defined as a 4×4 matrix [1 0 0 0; 0 1 0 0; 0 0 1 0; cmoon(1) cmoon(2) cmoon(3) 1] (with cmoon, the coordinates of the center of the Moon at a given elevation), and used for perspective transforms and 3-dimensional perspective division of the affine space encompassed by the sphere at infinity (inducing the 3-dimensional perspectival presentation of the world model in the FoC), so that the frame of the affine space was defined as $F_{affine}$ = $T_{persp}$*$F_0$.

Free energy ($Fe$) was minimized as a function of $\sigma$ across the range of $\lambda$ and $\mu$ to derive the optimal $argmin$($Fe[\lambda, \mu]$) following equations [39-40] (see Supplementary Information §2), using equation [26] for the calculation of areas, integrating over the angular range [0 : 1.6 * 10$^{-4}$ : $\pi$/2]. The change of metric in the sphere at infinity as induced by the minimization of $Fe$ and variable Moon elevations was computed as follows.

$\Theta$ = atan2($\lambda$*sqrt(sin($\phi$)$^2$ + $\mu^{(-2)}$*cos($\phi$)$^2$)*sin($\theta$), cos($\theta$)),
$\Phi$ = atan2($\mu$*sin($\phi$), cos($\phi$)),

with $\theta$ the elevation and $\phi$ the azimuth of the Moon (in radians), sqrt(.), the square root function, and atan2(.,.), the multi-valued inverse tangent function. The results were then expressed in homogeneous coordinates at infinity, so that:

(X,Y,Z,0) = [cos($\Theta$), sin($\Theta$)*cos($\Phi$), sin($\Theta$)*sin($\Phi$), 0].

The absolute area of the Moon in the FoC projective space was computed as: $\pi$*($\Delta$Z/2)*($\Delta$Y/2), with $\Delta$Z, the vertical elliptical length and $\Delta$Y, the horizontal elliptical length of the projected Moon. The area was then expressed as a relative area by dividing $area$($\theta$) at a given elevation by $area$($\theta$ = 10) at elevation 10 (near the horizon, and corresponding to the lowest elevation for ratings in the Virtual Reality experiment).

Rendering of the space for Fig. 2 was performed using Matlab's patch function.

**VR experiment**

**Participants**

There were N = 6 participants (Females = 4; age range [24 - 40]), with normal or corrected to normal vision. Informed consents were obtained following local IRB guidelines.

**VR setup**

The experiment was programmed in *Unity3d* version 2017.2.0f3. A scene was created with basic assets (ground plane, cubicles, road planes, mountains, moons) (Fig. 3), in order to manipulate perspective cues, based on stereoscopic vision through the Head Mounted

Display (HMD), geometrical perspective cues from objects due to their distance, and atmospheric effects (fog). Moon spheres, all of the same diameter (2,500 meter), were placed in the scenes at various elevation and azimuth, but at a constant distance from the location of the participants of 50,000 meters. The experiment used an HTC Vive system for the VR immersion in the scene. Scripts were programmed in C# to control conditions, the positions of moons, ratings, saving the results in a CSV file.

**Experimental conditions**

The following conditions were manipulated. Moon elevation, ranged from 2° (reference moon) to 10 to 70° (target moons), by steps of 10°, at constant azimuth of 90° with respect to x-axis in front). Terrain visibility 1: a) mountains and buildings with horizontal plane on; b) all terrain cues off (giving the impression of floating in space). We predicted that the Moon Illusion would be maximum in (a) and absent in (b). In order to assess possible elliptical effects in dedicated blocks, an array of additional lateral moons ranging from 2°to 20° in elevation and azimuth were used to maximize the likelihood of elliptical effects due to the complexity of frame optimization in such a context.

**Experimental schedule, tasks and ratings**

Conditions were controlled manually by dedicated key press performed by the experimenter following a pseudo-random schedule. Two experimental blocks were performed.
A) Manipulation: presence of terrain and perspective cues, condition (a) versus (b) above. For each condition and each trial, a target moon was presented at a given elevation, in sequences (from maximum to minimum, and minimum to maximum, ranging from 10° to 70°, among 7 possible basic elevations, repeated twice per participant), yielding 28 values × 2 conditions = 56 trials.
B) Manipulation: presence of terrain and perspective cues, condition (a) versus (b) above, with in all cases an array of additional moons present (see above). For each condition and each trial, a target moon was presented at a given elevation, in sequences (from maximum to minimum, and minimum to maximum, ranging from 10° to 70°, among 7 possible basic elevations, repeated twice per participant), yielding 28 values × 2 conditions = 56 trials.
Tasks and ratings were as follows. No time pressure was imposed. The total duration of the experiment, including setup, training and test was an hour on average.
In block A, participants were asked to perform the following. They used the two arrow keys (left-right) on the keyboard to decrease or increase the apparent overall diameter of a "reference moon" always presented at a 2° elevation above the horizon in front. The task was to make the apparent diameter of the reference moon match as closely as possible the apparent diameter of the "target moon" appearing at different elevations above ground. Participants were asked to look carefully at each moon as directly as possible: reference and target, alternatively, looking up or down, to perform the matching. Once satisfied, they indicated it verbally in order to move to the next trial.
In block B, participants were asked to perform the following. They used the two arrow keys (up-down) on the keyboard to decrease or increase the lateral diameter of the reference moon (this results in a shape of the moon that is more or less elliptical: from a vertical elongation if the diameter is decreased to a horizontal elongation if it is increased, with a perfectly round shape in between). The task was to make the apparent shape of the reference moon (irrespective of overall size) match as closely as possible that of the target moon.

Participants were asked to look carefully at each moon directly: reference and target, alternatively, looking up or down, to perform the matching. Once satisfied, they indicated it verbally in order to move to the next trial.

The size of the reference moon was reset between each trial.

**Procedure**

The experimenter set up and helped the participant get equipped with the requisite technology. The participant was instructed to sit on a rocking chair in the center of the room. The participant was given a keyboard to set on his/her legs in order to provide responses. The participant equipped with the HMD entered the virtual scene and was given a succinct introduction about Virtual Reality. Block A was followed by block B. There was a 2-minute pause in the middle of each experimental block and a 2-minute pause between blocks. During the pauses the participant was instructed to remove the HMD. At the end of the experiment, the experimenter helped the participant to remove all of the experimental equipment and provided a debriefing about the specific goals of the study.

**Data analysis**

All analyses were performed in Matlab (MathWorks$^{TM}$). The area of the apparent reference moon disks as a function of elevation $\theta$ was calculated as $A(\theta) = \pi * D(X) * D(Y)$, with $D(.)$ the diameter (in meters) of the moon. The areas were normalized by that of the target moon at 10° in order to express the areas as a proportion of the lower target moon area, so that $A(\theta) = A(\theta)/A(\theta = 10°)$. The elliptical bias (for block B) was computed as $abs(D(Y)/D(X))$. Data were averaged within participants for each elevation (2 samples per condition per elevation). Simulations were performed in order to regress empirical data on data predicted by the model. After considering the empirical range of relative areas (with minimum relative areas of around 0.6), the space of parameters of the PCM used to generate predicted curves for fitting, ranged as follow: $\sigma = [0.1 : 0.128 : 1.25]$; $C' = [0.001 : 0.006 : 0.6]$; C was fixed at C = 3. We thus varied two of the parameters of the model. Individual empirical data were first fitted on simulated data using a least square procedure. Simulated curves with least square were retained as best match for empirical data. Empirical data and simulated data were averaged across participants for each elevation and condition in order to estimate central tendencies. The fit of the average empirical and simulated data was then computed using linear regression in order to compute statistics and assess goodness of fit. Likewise, we used a linear regression to fit straight lines (both slope and intercept) to the average empirical data. We then compared the goodness of fit of the average empirical data with i) averaged simulation data from the PCM, and ii) the straight linear model (LM). We hypothesized that the PCM simulated data would demonstrate a better goodness of fit with the empirical data than the linear model. For statistical analyses, the averaged data and elevations, as well as the curves predicted by the PCM, were z-scored so as to enable a valid comparison across goodness of fit between the PCM and the linear models. The fitting of the empirical data on the simulated data also allowed us to estimate, in each individual and on average, the parameters $\lambda$, $\mu$, $\sigma$, and $C'$, corresponding to the projective frames of the FoC implied by the model, given the observed behavior of the relative areas as a function of elevation and condition. Goodness of fit was compared using the following indices: relative $R^2 = 1 - (MSE(PCM)/MSE(LM))$ (with MSE(.), the Mean Squared Error); AIC = $n*\log(MSE) + 2*k$ (with n the number of elevations, and k the numbers of parameters used for model optimization and fitting), for both the PCM and LM; BIC = $2*\log(MSE) + k*\log(n)$. k was equal to 3 for the PCM (2 parameters ($\sigma$ and $C'$), and 1 regression coefficient), and to 2 for the LM (1 regression coefficient, 1 intercept). We used a paired t-test to compare elliptical bias effects over the participants that performed the task. We report these results as initial results in support of the theory, but plan to perform a larger and more detailed empirical study in VR for a future report.

Note that, in the model, λ larger than 1 implies the displacement of the point I in the direction of H, and μ larger than 1 the displacement of I in the lateral direction. If one of them is smaller than 1, the displacement goes in the opposite direction.

**Author Contributions**

D.R. developed the study concept.
D.R., D.B. formulated the hypotheses.
D.B. developed the mathematical framework.
D.R. implemented the simulations and designed and conducted the VR study.
K.W. performed in-depth background analysis.
D.R., D.B., K.W. developed the rationale and discussion and wrote the manuscript.

**Acknowledgements**

We thank Karl Friston, Björn Merker, and Fabio Solari for their reviews during the preparation of this report.

**Supplementary Information**

**1. Projective Consciousness Model General Summary**

**General definitions**

**Active inference.** A formal and computational framework for modeling autonomous, embodied cognition, inspired by phenomenological traditions from Merleau-Ponty (*29*) to Varela (*30*) and formalized by Friston (*3*). Active inference is a method of information processing related to Bayesian inference by which an autonomous system: (i) anticipates the consequences of its actions by predicting how they will be experienced; (ii) programs its actions and acts accordingly, and (iii) updates its prior beliefs based on a comparison of its predictions and sensory evidence. According to the PCM, consciousness is governed by active inference. Agents run through cycles of perception, imagination, and action in order to optimize the precision of their knowledge and the satisfaction of their preferences.

**Free energy**. A quantity introduced in Statistical Physics and transposed into Information Theory and Bayesian Learning Theory, in which it is based on predictive coding and can be applied to active inference. The quantity is the sum of three terms: discrepancy between expectation and actual state, departure from prior beliefs, and negentropy. The free energy principle entails that agents attempt to minimize their overall free energy in order to maximize the validity of their expectations and the satisfaction of their preferences in a globally optimal manner. Formally, it is an upper bound on surprise. According to the PCM, consciousness minimizes free energy either factually or by anticipation, through its cycles of perception, imagination, and action, as well as prior updates. (Cf. below, Appendix 2.2.)

**Entropy**. A quantity, derived from Statistical Physics and forming the basis of Information Theory that characterizes the level of uncertainty about anticipated outcomes. Entropy is maximal for predictions with maximally uncertain outcomes and equal to 0 for completely certain ones. It scales up with the number of alternative possibilities. In the PCM, the more complex the inferences are, i.e., the more options and uncertainty, from simple physical situations to complex social interactions, the bigger the entropy to be processed as part of the optimization mechanism.

**Projective transformations**. Geometrical operations for transforming projective coordinates, extending the usual coordinates in Euclidean space. (In this paragraph, two kinds of transformations are considered together: the bijective ones, forming a group, that we used in this report for changing frames, and projections from one subspace to another, which are singular in 3-dimensions, and do not form a group.) Every non-zero linear transformation of a 4-dimensional vector space induces a projective transformation in the generalized sense, which is not defined on the projective subspace corresponding to the vectors cancelled out by the linear transformation (called its "kernel"). Such transformations can place points along directions of perspective in relation to a horizon at infinity and an implicit point of observation. They cover first-person and third-person perspectives. In the PCM, the changes of projective frame correspond to one-to-one projective transformations and implement spatial intentionality, attention, and perspective taking across perception, imagination, and action programming. They are selected based on the process of free energy minimization and reciprocally have impacts on perception and action coherence. (Cf. below, Appendix 2.1.)

**Field of consciousness (FoC).** The FoC is a (virtual) 3-dimensional projective space, and behaves in a manner that is analogous to force fields in physical theories, extended over space-time and internal variables and governed by the minimization of free energy. A fundamental property of the FoC is its reliance on projective changes of frame, altering global and local perceptions, the contents of consciousness, and thus influencing cognition and behavior. Its dynamics drive anticipation, orientation and action selection. The FoC provides a high degree of information integration across multiple sensory and cognitive modalities. It is organized around a "first-person point of view" but can simulate "other persons' points of view" and their relations, allowing the conscious organism to adaptively engage with its non-social and social surroundings. Note that the notion of the FoC has inter-related descriptive (phenomenological) components, and functional (biological, cognitive, affective, behavioral) components.

**Subjective rendering**. The set of (neuro)computational process whereby FoCs are generated and updated in response to new data (sensory, affective, semantic, etc.) or imaginary perspective taking. The subjective rendering engine relies on a dynamical process: the active inference based on free energy minimization modifies the states of internal variables; in particular it generates a projective geometrical transformation that alters information integration and influences conscious subjective perceptions and decisions, based on sensorimotor processing. It is called "subjective" rendering due to the oriented, "first-person point of view" structure inherent in every FoC and its phenomenal manifestation.

**Overall model description**

The Projective Consciousness Model, posits that the structure of conscious perceptual (and imaginary) space approximates a projective 3-space (e.g., $RP^3$, the projective space in 3-dimensions over the field of real numbers) and relies on the corresponding group of transformations (e.g., PLG(4), the projective linear group in 4-dimensions) to simulate and evaluate possible paths through, and efficiently navigate, the organism's environment. Path selection is driven by an active inference engine that aims at free energy (FE) minimization, roughly the minimization of surprise relative to the organism's prior beliefs, preferences, and ongoing sensory inputs. Many cases of perceptual ambiguity and multistability (e.g., the Necker Cube), can be explained, according to the PCM, in terms of the lack of perceptual cues sufficient to determine, in accordance with FE minimization, a single, canonical perspective or angle of view on the object in question, thus allowing for *ad libitum* oscillation between one orientation and another, one's transient arbitrary preference being the only remaining determinant (see (*1*)).

The projective Field of Consciousness (FoC) plays a central integrative role as part of a general process of active inference, guiding behavior via predictions about the likely sensory consequences of actions and updating in a Bayesian way in response to sensory feedback. The process maximizes the reliability of prior beliefs and the satisfaction of preferences, which are encoded as conditional probabilities. FE minimization and invariance maximization together define the choice of projective parameters used for perspective taking in perception and imagination (e.g., first-person versus third-person perspective taking).

To illustrate, consider a case of perceptual ambiguity occurring in a rabbit hunt. Imagine that the rabbit hunter cannot decide from his current vantage point if the animal in the bushes some ten meters away is a rabbit or a cat. In order to resolve the ambiguity, the hunter must decide what new visual perspective on the animal is required. In order to do that, he must be able to imagine (actively or passively) a number of accessible visual points of view, evaluate the advantages and disadvantages of these possible points of view, and select the best (or one of the best) to attempt to realize.  Here the choice will be determined by how conducive the imagined point of view would be (if realized) to the attainment of the hunter's goals (i.e., determining the identity of the animal and enabling a clear shot) and how difficult the point of view is to realize given these goals (e.g., Will attempting to realize it scare the animal off?).  The process of relating his current point of view on the situation to the imagined points of view is what we call "perspective taking".  And the process of selecting an optimal perspective in a given situation is driven by the general directive to satisfy preferences given the constraints provided by prior beliefs and sensory evidence (prior beliefs being subject to Bayesian updating in the light of new sensory evidence).  The hunter selects and enacts the realizable perspective (from among those he can envision) with the greatest probability of reducing his FE further.  Imagining those possible perspectives requires an implicit mastery of projective geometrical transformations; choosing a single one to actualize requires a selection regime, one driven by FE minimization, according to the PCM.

The projective structure of the space of conscious experience ("conscious space" for short) is most notable in visual perception, with its points of view, horizon lines, vanishing points where parallel lines seem to converge, and scaling effects, revealing the relative distances of familiar objects.  But this projective structure is by no means restricted to vision; in fact, it is a pervasive feature of multimodal, conscious 3D space. Arrows of direction in space and exchanges of points of view, are essential for the control of reaching, grip, locomotion, imagined displacement and social perspective taking (see (*1,2*)). It is not too surprising that the projective plane was first discovered and studied with the aim of explicating and systematizing the rules governing the depiction of 3D visual scenes on 2D surfaces.  It may at first appear somewhat more surprising that multimodal conscious space can itself be modelled in terms of projective 3-space.

It is due to this projective arrangement that conscious experience is always perspectival and capable of making basic spatial distinctions (here vs. there, closer, farther away, above, below, behind, etc.), imbued with an elusive but roughly localizable origin, which is not itself a distinguishable object within the space, and that it is capable of implicit and explicit perspectival imagination, able to infer how physical objects would look from various angles, distances, etc. Projective imagination, in turn, is also crucial for intersubjective understanding, empathy, and Theory of Mind.

In the PCM, sensorimotor data integration is dependent on the choice of a projective frame $F=T(F_0)$ on the 3-dimensional projective space, which completes the usual ambient affine space E by adding points at infinity, forming a plane at the horizon. This projective frame is chosen based on FE minimization, in agreement with data and prior probabilities, including valuations related to internal preferences and motivational parameters, and used for *optimal subjective rendering* in the FoC.

Furthermore, beyond its general framing, the representational contents of the FoC (animate and inanimate objects, structures and properties of physical objects, etc.) are determined on the basis of information encoded in memory and extracted from sensory evidence in order to build a world model *R* that is projected in perspective using the projective transformations *T* for subjective rendering in the FoC. The application of a projective transformation to the world model results in a conscious world model S. The transformations distribute spatial structures and motions in the FoC, as well as their associated FE, according to a point of view, direction of aim, and a scope that can be used for the evaluation of action; they also require calibration for perceptual inference.

The distribution of subjective information in the projective workspace is interpreted as given by the operation of T on R:

$$S(x', y', z', w) = T \cdot R(x, y, z, 1) \quad [1]$$

where (x,y,z) are affine coordinates in the frame F, and (x', y', z',w) are homogeneous coordinates in the frame T(F). The model is projected in the FoC 3-dimensional workspace in a first-person perspective (1PP) mode through perspective divide across the three spatial ambient dimensions *(x,y,z)* as:

$$S(1PP) = S(\frac{x'}{w}, \frac{y'}{w}, \frac{z'}{w}, 1) \quad [2]$$

The parameters of *T(t)* and *R(t)* are selected at a given time instant *t* based on a combination of sensory evidence and prior beliefs such that:

$$(T(t), R(t)) = argmin(FE(T, R)) \quad [3]$$

## 2. Mathematical framework

### a) Projective frames and metric transformations at infinity

Let there be given a real vector space W of dimension n+1. The projective space P(W) is the set of lines through 0 in W; it is said to have n dimensions. Every invertible linear transformation u from W to itself induces a transformation T of P(W) that is called a projective transformation. If v is a constant multiple of u, it defines the same T, and reciprocally if u and v define the same T, v is a constant multiple of u.

By definition, a projective subspace Q of dimension m in P(W) is a subset of the form P(U), where U is a linear subspace of W of dimension m+1. In particular, a hyperplane of P(W) is a subset P(V) where V is a linear subspace of dimension n, i.e., co-dimension 1 in W.

A linear basis $e_1,\ldots,e_{n+1}$ of W defines homogeneous coordinates $[x,\ldots,x_{n+1}]$ on P(W) that are n+1 numbers, not all equal to zero, considered to define the point in P(W) (i.e., the same line in W) which passes through the vector $x = x_1 e_1 + \ldots + x_n + 1 e_{n+1}$. Thus two sets of coordinates $[x_1, \ldots, x_{n+1}]$ and $[x'_1,\ldots,x'_{n+1}]$ are equivalent if and only if there exists a non-zero constant c such that $x'_1 = cx_1,\ldots,x'_{n+1} = cx_{n+1}$.

**Definition**: A *projective frame* in P(W) is a set of n+2 (projective) points $P_0, P_1, \ldots, P_{n+1}$, such that no subset with n+1 elements belongs to a hyperplane of P(W).

**Proposition 1**: In this case there exists a linear basis $e_1,\ldots,e_{n+1}$ of W, such that $P_j$ corresponds to the line generated by $e_j$ for any j between 1 and n+1, and $P_0$ corresponds to the line generated generated by $e_0 = e_1 + \ldots + e_{n+1}$. Moreover, this basis is unique up to the multiplication of all vectors by a common constant.

*Demonstration*: We choose a basis $e'_1,\ldots,e'_{n+1}$ of W such that $e'_j$ generates the line $P_j$, then the numbers $x'_1,\ldots,x'_{n+1}$ such that $x'_1 e'_1 + \ldots + x'_n + 1 e'_{n+1}$ generates $P_0$ are well defined up to the multiplication of all of them by a common constant. The searched frame is given by putting $e_j = x'_j e'_j$ for any j between 1 and n+1.

From this proposition, it follows that a projective frame is equivalent to n+1 vectors linearly independent up to multiplication by a non-zero constant. However, this fact is a bit misleading in view of the following result, which shows that n+1 points are truly not sufficient for making a projective basis.

**Theorem 1**: Given two projective frames $P_0,\ldots,P_{n+1}$ and $P'_0,\ldots,P'_{n+1}$, there exists a unique projective transformation T such that for every j, $T(P_j)=P'_j$.

*Demonstration*: We choose two linear bases of W, $e_j$ and $e'_j$ ($j=1,\ldots,n=1$) that are related to these two frames as described in the statement of the preceding proposition (not its proof), then there exists a unique linear application u such that $u(e_j)=e'_j$ (for $j=1,\ldots,n=1$). The relation $u(e_0)=e'_0$ is automatic by linearity. This defines T.

Historically, projective spaces were defined by adding points at infinity to affine spaces. In practice, it is important to work in such a context, and thus to explain how it relates to what precedes just above.

Let V be any hyperplane in W, defined by a linear equation $a(w)=0$. Then the subset E of W defined by $a(w)=1$, has a well-defined structure of affine space associated to V: given two points A and B in E, there exists one and only one vector v in V such that $B=A+v$. The points P of E can be identified with points of P(W) by taking the lines containing 0 and P; those projective points are named the "points at finite distance" of P(W), even if there exists no preferred distance here. (In fact, a Euclidean metric on E is equivalent to a positive non-degenerate scalar product on V.) And the projective subspace P(V) of P(W) can be seen as the hyperplane at infinity of E, by associating the directions in E parallel to non-zero vectors in V. An affine frame of E is made by a point $P_{n+1}$ and a linear basis $e_1,\ldots,e_n$ of V. Given such an affine frame, an associated linear basis of W follows by taking for $P_j$ the lines generated by $e_j$ when $j=1,\ldots,n$, and taking $e_{n+1}=P_{n+1}$. Then we deduce a projective frame, just by taking the point $P_0$ as before, i.e., the line generated by the vector $e_0=e_1+\ldots+e_n+e_{n+1}$.

Conversely, if a projective frame $P_0, P_1,\ldots,P_{n+1}$ is given, according to *Proposition 1*, this defines a unique basis $e_1,\ldots,e_{n+1}$ of W. We define V as the vector subspace of W generated by $e_1,\ldots,e_n$, equipped with the equation $a(v)=0$, such that $a(e_{n+1})=1$. This defines an affine space E as before. The points $P_1,\ldots, P_n$ belong to the plane at infinity P(V) which completes E to get P(W); in E we find $P_{n+1}$ corresponding to the line generated by $e_{n+1}$ and $P_0$ corresponding to the line generated by the sum $e_0=e_1+\ldots+e_{n+1}$. The intersection of the line $P_{n+1}P_0$ with P(V) is the point $Q_0$ that corresponds to the vector $f_0=e_1+\ldots+e_n=e_0-e_{n+1}$, because this is the unique vector such that $P_0=P_{n+1}+f_0$ belongs to the affine space E.

**Remark**: It is a pure convention to choose this form of $e_0$; for instance, nothing essential would be changed if we had chosen $e'_0=-e_1+\ldots+e_{n+1}$. This could be seen as another frame $P'_0,P_1,\ldots,P_{n+1}$, and this would have no effect on the projective transformations associated with another frame, on the condition that we respect the form of the combination of basis vectors, the same sign at the same place. This remark points to the important fact that there exists a manifold of concrete representations of what a frame is; this could, for example, be a point cloud, deduced one from another by fixed conventional changes of coordinates.

In the 3-dimensional projective space presented in the body of the text, the points L, V, H correspond respectively to $P_1, P_2, P_3$, the point $P_4$ to O, and the point $P_0$ to I. The point $Q_0$ is the point J. When modifying the convention as in the above remark, we exchange left and right, then the point cloud R/L and I, I' is an example of the possible extension of the notion of a frame.

Let us now look at metrics in the space P(V). Elliptical metrics (corresponding to spherical metrics of the two-fold cover of P(V) by the sphere of oriented directions in E) appear more natural, but we will also consider Euclidean metrics, describing an affine part of P(V) centered on $P_n$ (that correspond to our H at the horizon), because some subjects have the tendency to report their estimation preferentially in these Euclidean terms. In this Euclidean case, we assume that $P_1,..,P_{n-1}$ are at infinity in P(V) (corresponding to V and R/L). (Later in this supplementary document, in (c), we will also consider natural hyperbolic metrics on parts of the projective space.)

We admit that the standard elliptical (or spherical) metric that corresponds to the usual measure of angles, is associated with a standard frame of reference (that we named the frame by default ($F_0$) in the main text), the largest angle is $\pi/2$, it is realized for any pair $P_j$, $P_k$ with j,k between 1 and n, and distinct. Then the angle between $Q_0$ and a $P_j$ is $\pi/4$.

Let us look at the effect of changing the frame into a new one $P'_0,\ldots,P'_{n+1}$. We restrict ourselves in what follows (except in (c)) to the case where $P'_1,\ldots,P'_n$ continue to belong to the hyperplane at infinity P(V). There exists a unique projective isomorphism T of P(W) sending a $P_j$ to the corresponding $P'_j$, for j=0,1,…,n+1. There exists a unique rescaling of the corresponding linear isomorphism u of W (a 4×4 matrix, in our case n=3) such that u preserves the affine part E, and induces an affine transformation on it, which can be identified with the restriction $T_E$ of T.

Let us look first at the case where T fixes all the points $P_j$, j=1,…,n+1, and moves only $P_0$ into a point $P'_0$, then $Q_0$ into $Q'_0$ in P(V) (in our 3D case, this means that only I and J are moved). The new basis $e'_1,\ldots,e'_n$, $e'_{n+1}$ of W, is made of proportional vectors, $e'_1=a'_1 e_1$, …, $e'_{n+1}=a'_n+1 e_{n+1}$, i.e., the matrix of u in the standard basis $e_1,\ldots,e_n$ is diagonal with entries $a'_1,\ldots,a'_{n+1}$. Due to the fact that u preserves E, the last entry $a'_{n+1}$ is equal to 1, but the other ones can be anything except zero, and can be interpreted as the coordinates of P'0 in the affine frame of E centered in $P_{n+1}$=O, and with axis normalized by $e_1,\ldots,e_n$, that is because $e'_0=e'_1+\ldots+e'_{n+1}=e_{n+1}+a'_1 e_1+\ldots+a'_n e_n$, generates the line $P'_0$ and cuts the affine hyperplane E of W in the point $O+a'_1 e_1+\ldots+a'_n e_n$. Therefore $Q'_0$ is defined by the vector $e'_0-e_{n+1}=a'_1 e_1+\ldots a'_n e_n$, which gives another interpretation of the coefficients $a'_1,\ldots,a'_n$. For $Q'_0$, and for the corresponding transformation in P(V), and then for the angular geometry, only the homogeneous coordinates $[a'_1,\ldots,a'_n]$ play a role, i.e., we can multiply all of them by the same non-zero number without changing this geometry. (If we were interested in the geometry at finite distance, on E, we should have considered the separate values of all the coordinates.)

For every j and k between 1 and n, the arc from $P_j$ to $P_k$ continues to have the value $\pi/2$, and the arc from $P_j$ to $Q'_0$ now is equal to $\pi/4$.

The easiest way to express the change of geometry induced by the new point $Q'_0$, replacing $Q_0$, is to consider the ellipsoid Σ' in E of equation ${a'_1}^2 x_1^2 + \cdots + {a'_n}^2 x_n^2 = 1$. Its half-axis in the direction of $e_j$, j=1,…,n, is $1/a'_j$. This ellipsoid is the unit sphere in the new coordinates $x'_j$, j=1,…,n. Let us denote by Σ the unit sphere in the old coordinates $x_j$, j=1,…,n, i.e. $x_1^2 + \cdots + x_n^2 = 1$. And consider the radial projection ρ from Σ to Σ', followed by the inverse of the transformation u, to come back to Σ; this composed transformation f gives the new metric on Σ. In terms of homogeneous coordinates f is nothing but the inverse $T^{-1}$ of T.

For example, if n=2, let us choose $a'_1=\lambda$ and $a'_2=1$, and denote the coordinates $x_1$ and $x_2$ by the letters x and y respectively. The sphere Σ is the ordinary circle of radius one and center 0, $x^2+y^2=1$, and the ellipsoid Σ' is the ellipse of equation $\lambda^2 x^2 + y^2 = 1$. Let us parametrize Σ by the angle θ=arctan(y/x), then ρ(θ)=($\lambda^{-1}/\sqrt{1+\lambda^{-2}tan(\theta)^2}, \lambda^{-1}\tan(\theta)/\sqrt{1+\lambda^{-2}tan(\theta)^2}$), and $f(\theta) = u^{-1}(\rho(\theta)) = \arctan(\lambda\tan\theta)$. To understand the deformation of the metric on the circle Σ that is induced by f, we have to compute its derivative at a given elevation θ, this is f'(θ)=$\frac{\lambda\tan'(\theta)}{1+\lambda^2 tan(\theta)^2} = \lambda\frac{1+tan(\theta)^2}{1+\lambda^2 tan(\theta)^2} = \lambda\frac{1}{cos(\theta)^2+\lambda^2 sin(\theta)^2}$. We see that for zero elevation the factor of expansion (or contraction if λ<

1) is the number λ, and for elevation at 45° it is equal to 2λ/(1+$\lambda^2$). To go further we compute the second derivative:

$$f''(\theta) = \lambda(1-\lambda^2)\frac{2\cos(\theta)\sin(\theta)}{(\cos(\theta)^2+\lambda^2\sin(\theta)^2)^2} \quad [4]$$

Thus f is a concave function when λ is strictly larger than 1, which means that the factor of expansion f'(θ) is larger and larger when θ approaches 0; we referred to this as the convexity of the Moon dilatation when it approaches the horizon.

For n=3, we parametrize the round sphere Σ by the longitude θ and the latitude φ, in such a manner that x=$x_1$=cos θ, y=$x_2$=sin θ cos φ and z=sin θ sin φ, the transformation u being given by x'=λ x, y'=μ y and z'=z. (With the notations in the body of the text x is along OH, y along OR, and z along OV. The coordinates of I are noted (λ,μ,ν).)

Applying ρ and the inverse of u, we obtain the following transformation f of the sphere Σ: f(x,y,z)=(X,Y,Z); where: X=$\lambda^{-1}$cos (θ)/R ; Y=$\mu^{-1}$sin(θ)cos(φ)/R; Z= $\nu^{-1}$sin(θ)sin(φ)/R, and, R=$\sqrt{\nu^{-2}\sin(\varphi)^2\sin(\theta)^2 + \lambda^{-2}\cos(\theta)^2 + \mu^{-2}\cos(\varphi)^2\sin(\theta)^2}$.

In spherical coordinates (Θ, Φ), this yields:

Θ=arctan(λ$\sqrt{\nu^{-2}\sin(\varphi)^2 + \mu^{-2}\cos(\varphi)^2}$tan(θ))  [5]
Φ=arctan(μ$\nu^{-1}$tan(φ))    [6]

In natural homogeneous coordinates on the plane at infinity, the transformation f is simply:

f(x ;y ;z)=(x/λ; y/μ; z/ν)   [7]

In the affine coordinates (y,z) centred in H in P(V), y lateral and z vertical, it is:

$f_a$(y ;z)=(λy/μ; λz/ν)    [8]

Then if the Euclidean structure in these coordinates is used, the metric is dilated if λ is larger than 1, and more dilated in the vertical direction than in the lateral one if μ is larger than ν, which might play a role under some conditions in the reported egg shaped of the halos around the Moon at the horizon (9).

Since to our knowledge an egg shape has not been reported for the Moon itself, we could assume a strong prior, consistent with visual angles, favoring the preservation of a round shape at infinity. Or perhaps the effect is too slight to be easily noticed.

Transformations that send circles to circles (without necessarily sending the center to the center) are called *conformal*. They are the transformations that multiply the metric tensor by a strictly positive function. We will see in Proposition 3 below that the only conformal maps of a projective plane are isometries, as opposed to what happens for the sphere.

We remark that the affine model $f_a$ is conformal when μ=ν, but the correspondence between the hemisphere x positive, or the projective plane, with the affine plane, which is given by central projection (called the gnomonic projection), is not a conformal map, as opposed to the projection of the full sphere on the tangent plane at a pole from the opposite pole (called the stereographic projection).

Let us compute how far the map f defined above is from a conformal map. The elliptical (or spherical) element of length ds is given by:

$$ds^2 = d\theta^2 + sin^2\theta d\varphi^2 \qquad [9]$$

Let us define $\rho = \sqrt{\mu^2 sin(\varphi)^2 + v^2 cos(\varphi)^2}$. We have $\frac{d\Theta}{d\theta} = \frac{\lambda\rho\mu v}{\lambda^2\rho^2 sin(\theta)^2 + \mu^2 v^2 cos(\theta)^2}$, $\frac{d\Phi}{d\varphi} = \frac{\mu v}{\mu^2 sin(\varphi)^2 + v^2 cos(\varphi)^2}$, and $sin\Theta = \frac{\lambda\rho sin\theta}{\sqrt{\lambda^2\rho^2 sin(\theta)^2 + \mu^2 v^2 cos(\theta)^2}}$. Therefore :

$$d\Theta^2 + sin^2\Theta d\Phi^2 = \frac{\mu^2 v^2(\lambda^2\rho^6 d\theta^2 + \lambda^2\rho^2 sin^2\theta(\lambda^2\rho^2 sin(\theta)^2 + \mu^2 v^2 cos(\theta)^2)d\varphi^2)}{((\lambda^2\rho^2 sin(\theta)^2 + \mu^2 v^2 cos(\theta)^2)(\mu^2 sin(\varphi)^2 + v^2 cos(\varphi)^2))^2}$$

$$= \frac{\mu^2 v^2 \lambda^2 \rho^2 (d\theta^2 + sin^2\theta d\varphi^2) + \mu^2 v^2 \lambda^2 \rho^{-2}(sin^2\theta(\lambda^2\rho^2 sin(\theta)^2 + \mu^2 v^2 cos(\theta)^2) - \rho^4 sin^2\theta)d\varphi^2}{(\lambda^2\rho^2 sin(\theta)^2 + \mu^2 v^2 cos(\theta)^2)^2} \qquad [10]$$

Thus the departure from conformality is given by:

$$\varepsilon d\varphi^2 = \frac{\mu^2 v^2 \lambda^2 \rho^{-2} sin^2\theta((\lambda^2\rho^2 sin(\theta)^2 + \mu^2 v^2 cos(\theta)^2) - \rho^4) d\varphi^2}{(\lambda^2\rho^2 sin(\theta)^2 + cos(\theta)^2)^2} \qquad [11]$$

More explicitly:

$$(\lambda^2\rho^2 sin(\theta)^2 + \mu^2 v^2 cos(\theta)^2) - \rho^4 = \mu^2\lambda^2 sin(\theta)^2 sin(\varphi)^2 + v^2\lambda^2 sin(\theta)^2 cos(\varphi)^2 + \mu^2 v^2 cos(\theta)^2 - \mu^4 sin(\varphi)^4 - v^4 cos(\varphi)^4 - 2\mu^2 v^2 sin(\varphi)^2 cos(\varphi)^2 \qquad [12]$$

For instance, on the vertical axis, where φ=π/2, the departure from conformality is zero if and only if :

$$\mu = \sqrt{\lambda^2 sin(\theta)^2 + v^2 cos(\theta)^2} \qquad [13]$$

Consequently, given the elevation θ of a point M along the vertical axis at infinity and the parameters λ, υ, there is a unique choice of μ preserving the circular form in the neighborhood of M. The same result holds true in any direction, because the following equation in $r = \rho^2$:

$$r^2 - \lambda^2 r sin(\theta)^2 - \mu^2 v^2 cos(\theta)^2 = 0 \qquad [14]$$

has one and only positive root for any value of θ strictly between –π/2 and π/2 and for any value m of the product μυ, which is given by:

$$r = \frac{\lambda^2 sin(\theta)^2 + \sqrt{\lambda^4 sin(\theta)^4 + 4m^2 cos(\theta)^2}}{2} \qquad [15]$$

which leads to the following equation in μ and υ:

$$2\mu^2 sin\varphi^2 + 2m^2\mu^{-2} cos\varphi^2 - \lambda^2 sin(\theta)^2 = \sqrt{\lambda^4 sin(\theta)^4 + 4m^2 cos(\theta)^2} \qquad [16]$$

By multiplication by the square t of μ, we obtain a second degree equation in t, with a strictly positive discriminant as soon as θ is smaller than π/4, due to the inequality:

$$4m^2 cos(\theta)^2 > 8m^2 sin\varphi^2 cos(\varphi)^2 \qquad [17]$$

There exists one and only one strictly positive solution of this equation, due to the minus sign of the term of degree one. Thus we have proved the following result.

**Proposition 2:** For any direction φ around H, and any value of θ smaller than π/4, and any strictly positive value of λ, there exists a one parameter family of pairs (μ, υ) making the transformation f conformal in the point M of coordinates (θ, φ).

This has to be contrasted with the following negative result.

**Proposition 3:** Any projective conformal transformation of the projective plane is an isometry and thus induced by a rotation of the sphere; that is the two-fold cover of the projective plane.
*Proof:* any projective isomorphism of the plane can be lifted to a bijection of the two-dimensional Riemann sphere, which commutes with the antipodal map τ; by composing with an orthogonal reflexion we can suppose that this transformation preserves the orientation, then, if it is conformal, it is expressed by a complex homography, which has one or two fixed points. The case of a unique fixed point is excluded by the commutation with τ. Thus we have two antipodal fixed points. The eigenvalues of the linear approximations at these two points must be inverse imaginary numbers (the determinant is one), having the same modulus (from the commutation with τ). Thus the homography is a rotation.

We have to compute the area A(λ,μ,υ) of the image by f of the domain Ω where θ is smaller than π/4 and φ is arbitrary. It is given by standard elliptic integrals of the third kind.

$$A(\lambda, \mu, \upsilon) = \int_0^{\pi/4} d\theta \int_0^{\pi} d\varphi \frac{d\Phi}{d\varphi} \frac{\partial \Theta}{\partial \theta} \sin(\Theta) \qquad [18]$$

Then:

$$A(\lambda, \mu, \upsilon) = \int_0^{\pi/4} d\theta \int_0^{\pi} d\varphi \frac{\mu\upsilon}{\mu^2 sin(\varphi)^2 + \upsilon^2 cos(\varphi)^2} \frac{\lambda\rho\mu\upsilon}{\lambda^2 \rho^2 sin(\theta)^2 + \mu^2 \upsilon^2 cos(\theta)^2} \frac{\lambda\rho sin(\theta)}{\sqrt{\lambda^2 \rho^2 sin(\theta)^2 + \mu^2 \upsilon^2 cos(\theta)^2}} \qquad [19]$$

To integrate over θ, we use the following formula:

$$\frac{d}{d\theta} \frac{cos(\theta)}{\sqrt{\lambda^2 \rho^2 sin(\theta)^2 + \mu^2 \upsilon^2 cos(\theta)^2}} = \frac{-\lambda^2 \rho^2 sin(\theta)}{(\lambda^2 \rho^2 sin(\theta)^2 + \mu^2 \upsilon^2 cos(\theta)^2)\sqrt{\lambda^2 \rho^2 sin(\theta)^2 + \mu^2 \upsilon^2 cos(\theta)^2}} \qquad [20]$$

We get:

$$A(\lambda, \mu, \upsilon) = \int_0^{\pi} \frac{\mu\upsilon d\varphi}{\mu^2 sin(\varphi)^2 + \upsilon^2 cos(\varphi)^2} - \int_0^{\pi} \frac{\mu^2 \upsilon^2 d\varphi}{(\mu^2 sin(\varphi)^2 + \upsilon^2 cos(\varphi)^2)\sqrt{\lambda^2(\mu^2 sin(\varphi)^2 + \upsilon^2 cos(\varphi)^2) + \mu^2 \upsilon^2}} \qquad [21]$$

The first integral can be computed easily, by introducing the variable t=tan φ:

$$\int_0^{\pi} \frac{\mu\upsilon d\varphi}{\mu^2 sin(\varphi)^2 + \upsilon^2 cos(\varphi)^2} = 2\mu\upsilon \int_0^{+\infty} \frac{dt}{\mu^2 t^2 + \upsilon^2} = 2\int_0^{+\infty} \frac{du}{u^2 + 1} = \pi \qquad [22]$$

The second integral is a multiple of the standard complete elliptic integral of the third kind in the form of Legendre:

$$\Pi(n,k) = \int_0^{\pi/2} \frac{d\varphi}{(1-n\sin(\varphi)^2)\sqrt{1-k^2\sin(\varphi)^2}} \quad [23]$$

with :

$$n = 1 - \mu^2 \upsilon^{-2} \quad [24]$$

and :

$$k^2 = \frac{\lambda^2\upsilon^2 - \lambda^2\mu^2}{\lambda^2\upsilon^2 + \mu^2\upsilon^2} = \frac{\lambda^2}{\upsilon^2}\frac{\upsilon^2 - \mu^2}{\lambda^2 + \mu^2} \quad [25]$$

Then the exact formula of the area, with these notations is

$$A(\lambda,\mu,\upsilon) = \pi - 2\frac{\mu^2}{\upsilon\sqrt{\lambda^2+\mu^2}}\Pi(n,k) \quad [26]$$

Also here we neglect nothing that is essential except symmetry, by imposing υ=1, and varying λ and μ.

**b) Free energy and projective frames**

**General considerations**

In general, variational Bayesian learning and decision depend on the minimization of a functional FE, similar to a free energy in statistical Physics. The variable is an *a posteriori* probability *p* on the internal parameters representing the states of the world, the states of the mind and body, including notably motivations and intentions. All these variables being denoted by letters X, Y, etc. Thus functional FE expresses a trade-off between the conservation of the *a priori* probability $p_a$ on the internal parameters and the best possible explanation of the probability $p_L$ on a subset of the variables $X_L$ that in general represent new observations, but that could as well represent new goals or beliefs:

$$FE(p) = E_p(-\log(p_a) + D_{KL}(X_L * p, p_L)) - S(p) \quad [27]$$

The entropy, which measures the total uncertainty, is defined as:

$$S(p) = -\sum p(x)\log(p(x)) \quad [28]$$

and the Kullback-Leibler divergence, measuring statistical proximity, is defined as

$$D_{KL}(p1,p2) = \sum p1\log(p1/p2) \quad [29]$$

We do not justify this function here: it is traditionally deduced from the lower bound of a probability (cf. (*3*)), but it is also directly a Kullback-Leibler divergence between probabilities on the product of the space of parameters and the space of the values of all variables.

In our application, the data on $X_L$ is not treated as a probability, but as a fixed value $x_L$, then the term $E_p(D_{KL})$ (the expectation of the Kullback-Leibler divergence of two probabilities) is replaced by $-\log(p(X_L=x_L)$, which yields the following simplified form:

$$\text{FE}(p) = -\log(p(X_L = x_L)) + E_p(-\log(p_a)) - S(p) \qquad [30]$$

However, even the minimization of this function turns out to be very difficult, thus people (and probably their brains too) have adopted a simplified version. Such simplification could be implemented through regionalization, as introduced by Bethe and generalized by Kikuchi (*31*), or by introducing a virtual probability q on all the variables X, and using the Jensen's inequality of convexity, which simplifies the form of FE and yields a more tractable function :

$$\text{FE}(p) = -\log(p(X_L = x_L)) + E_p\left(-\log E_q\left(\frac{p_a(x,\theta)}{p(x,\theta)q(x)}\right)\right)$$

$$\leq -\log(p(X_L = x_L)) + E_p\left(E_q\left(-\log\left(\frac{p_a(x,\theta)}{p(x,\theta)q(x)}\right)\right)\right)$$

$$= -\log(p(X_L = x_L)) + E_p E_q(-\log(p_a(x,\theta))) - S(p) - S(q) \qquad [31]$$

This function is now considered as a function F(p,q) of the pair or probabilities (q,p).

A further simplification consists in assuming that on θ the marginalization of p gives a certainty, and that for this value of θ, p(x, θ) coincides with q(x), and $p_a(x,\theta)$ becomes $p_a(x)$, which yields another form of free energy:

$$Fe(q) = -\log(q(X_L = x_L)) + E_q(-\log(p_a(x))) - S(q) \qquad [32]$$

The variable q now belongs to a chosen set Q of prescribed probabilities on X. As X is a joint variable ($X_k$; k in K), the choice of Q implies some hypotheses on the dependency between the internal variables $X_k$. The set L is considered as a subset of K.

In the PCM, one of the variables is a projective frame $F=T(F_0)$ on the 3-dimensional projective space P(W), which completes the usual ambient affine space E by adding points at infinity. Sensorimotor integration relies in particular on F. We will assume for simplicity that the law q on F (or T) has a Gaussian shape with small variance. In fact, a variance zero would lead to a contradiction because the *a priori* –log(p) would take an infinite value in this case. Other variables are observables or hidden variables. The law on the projective frame, as the laws on the other variables, is chosen based on the minimization of FE. Once F is chosen, and the other elements of the law q, perceptions and actions follow, which change the data, so that a dynamical system emerges. We replace the prior $p_a$ by a probability p deduced from q; and a new F and new p are determined for the next time instant, and so on.

**Application to the Moon Illusion**

When the Moon appears in the sky, one of the principal inputs $x_L$ is its approximate position in the sky. Other inputs from $X_L$ are defined by the visible landscape and objects. In what follows, we adopt the notations of the first section of this appendix.

The prior distribution $p_a$ on the frame is centered in $F_0$ and has variance $\sigma_0$, which corresponds to a matrix, itself constituted by four 4×4 matrices, one for each of the four vectors $e_1,e_2,e_3,e_4$ in the 4-dimensional space W. This representation is simpler than the (non-equivalent) representation of variance using five 3×3 matrices for the points O, H, V, L, I.

For minimizing the second term in $Fe(q)$, the frame F has to depart as little as possible from $F_0$. In this case, the variation of H,V,R/L and O is minimal, but the variation of I has a cost. If we take the complete contribution of the variable F to the free energy function *Fe*, we get the $D_{KL}$ between two Gaussian laws in the vector space V: one for $f_0=e_1+e_2+e_3$, and the other for f'$_0$=a'$_1$e$_1$+a'$_2$e$_2$+a'$_3$e$_3$. This is known as the sum of squares of the Euclidean distance, corresponding to the covariance I*σ$_0$ of the first law induced on the fifth point I, between the points in 3D space and a non-symmetric function of the two covariance matrices.

In what follows we neglect this second part, but it could have an effect. Note, in addition, that visual information, which unconsciously puts the horizon farther, i.e., augments d, as in Kaufman & Kaufman (*12*), has the effect of displacing f'$_0$ in the direction of e$_1$ (i.e., H), thus augmenting λ and yielding a larger visible horizon, elongating the frame in the lateral direction e$_2$ (i.e., R/L), which increases μ. By symmetry, the only unknown parameter here is the scale of I*σ$_0$, the statistical unit σ: it is smaller than 1 if the prior in favor of the standard frame by default is strong, and is larger than 1 if this prior is weak, and:

$$Fe(F) = \frac{1}{2\sigma^2}((\lambda - 1)^2 + (\mu - 1)^2) \quad [33]$$

For the other terms of *Fe*, the role of F is indirect: we assume that choice of F has an effect on the quality and quantity of observations inside the cone delimited by I. Then the variables X$_L$ that can impact the choice of F are observables in the vicinity of the Moon, and between the Moon and the observer. If the Moon is full and luminous, and if salient cues are present, many variables can be estimated more accurately. Thus, for minimizing the first term in *Fe*, that is $-\log(q(X_L = x_L))$, the point I has to be chosen far in the direction of the moon when it is near horizon. Contrariwise, in the case of the Moon high in the sky, the point I is better placed more laterally in order to augment the likelihood of acquiring new information. The problem now is to quantify this difference.

Concerning all the variables X$_K$, including X$_L$, the two last terms in *Fe* taken together correspond to the KL divergence from fine graining q to coarse graining p$_a$. We can assume that most of the variables X$_L$ are indifferent for the prior, i.e., p$_a$ is uniform. In this case, it has a large entropy, and - log(p$_a$) is a constant U$_a$. Then E$_q$(-log(p$_a$))=U$_a$ too, and:

D$_{KL}$(q;p$_a$) =U$_a$-S(q).

D$_{KL}$ decreases if the entropy of q increases. This entropy is large if the quantity and quality of cues are large, and is small in the opposite case. This is because, entropy grows with the number of states that are estimated. We can assume that such quality, for each variable X$_L$, is proportional to the sum of the logarithms of the luminosity of the Moon, say γ, measured in units inverse to the area, of the area α of the visible object, of the quality of sight, say ω (two eyes and so on), and of the quality κ of the object (contrast and so on).

Let us denote Λ= log(αγ) + log(ωκ); we have:

$$S(q_L) = CN_L\Lambda \quad [34]$$

where C is a normalization constant, and N$_L$ is the number of observables for the sight in the region of interest.

The constants α, γ, ω and κ are independent of the choice of F, and can be manipulated experimentally. The quantity N$_L$, however, depends on F. It corresponds to the number of available "independent variables" in X', or a number of accessible dimensions, related to the available information. We hypothesize that N$_L$ is proportional to the inverse of the area of the region determined by J around H.

When μ=1, all the necessary formulas can be explicitly computed.

In the elliptical geometry at infinity, that is, half the standard spherical geometry, the area *a* of the region defined by θ≤θ$_0$ is a(θ$_0$)=π(1-cos(θ$_0$)).

Then the ratio of areas, between the region covered by J$_0$ and the region covered by J is given, according to the function f of section 1 (i.e., f(θ)=atan(λ tan(θ))), by the following formula as a function of λ:

$$\rho(\lambda)= (\sqrt{2} - 1)\sqrt{(1 + \lambda^2)}/\sqrt{2}(\sqrt{(1 + \lambda^2)} - 1) \qquad [35]$$

Then we want to minimize the following function:

$$Fe(\lambda)= \frac{1}{2\sigma^2}(\lambda - 1)^2 + C'\Lambda\rho(\lambda) \qquad [36]$$

where C' is a constant independent of everything else.

This function $Fe(\lambda)$ is strictly convex and tends to infinity when λ tends to infinity or zero; therefore, it has a unique minimum.

However, the prior preference assumed here based on phenomenology for keeping a Moon of round shape makes the intervention of lateral parameters μ and υ necessary. In what follows, to simplify, we fix υ=1, which is not a problem if the angle φ is sufficiently large, say between 30° and 120°, which is quite a natural hypothesis in our context, given the adopted coordinate system. In this case, we have seen that, given a point M(θ,φ), a unique μ exists such that the transformation f is approximately conformal in the vicinity of M. To simplify the simulations, we work with M on the vertical, thus with υ=1 and:

$$\mu = \sqrt{\lambda^2 sin(\theta)^2 + cos(\theta)^2} \qquad [37]$$

We will now denote by ω the value of θ where the Moon is seen.
We define :

$$\rho(\lambda, \mu) = \frac{1-1/\sqrt{2}}{A(\lambda,\mu,1)} \qquad [38]$$

where A is given by the complete elliptic integrals in section (a).

Thus two components of $Fe(\lambda,\mu)$ contribute to the overall FE, with a differential weight depending on the position of M with respect to a transition point for elevation.

When the elevation of M is under that point (say π/4 where in many cases no environmental information will appear in the visual field), the function that contributes the most to FE to be minimized is:

$$Fe(\lambda,\mu)=\frac{1}{2\sigma^2}\left((\lambda - 1)^2 + C(\omega)(\mu - \mu(\lambda, \omega))^2\right) + C'\Lambda\rho(\lambda,\mu) \qquad [39]$$

where C expresses the strength of the *a priori* assumption of roundness. We introduced a dependency on ω to take into account that the departure from roundness varies with the Moon's elevation center $\omega$. For small values of $\omega$, μ values yield roundness, but for larger ones, a precise tuning is warranted, leading to a larger weight $C(\omega)$, which can be defined as $C_0 sin^2(\omega)$, for respecting the order of the defect of conformality.

In general, in a model preserving the roundness of the Moon's shape (i.e., with a conformal behavior), C is supposed to be large as a prior. It is smaller if conformality is not encoded as a strong prior. The form of the function *Fe* implies that its minimum is to be found for λ and μ larger than 1. This generates focalization to the horizon HL, accompanied by a corrective relative enlargement for the lateral integration of information.

When the Moon is higher in the sky beyond the transition point (say more than 45° of elevation where only the open sky is given and no more proximal cues are present), according to our main rationale, the projective frame has to further shift from fine graining to coarse graining to maximize information integration. Thus the region of integration is expected to be further enlarged and the equivalent area to grow. Consequently, the function that contributes the most to FE to be minimized is:

$$Fe'(\lambda,\mu) = \frac{1}{2\sigma^2}\left((\lambda-1)^2 + C(\mu - \mu(\lambda,\omega))^2\right) + C'' \Lambda\, A(\lambda,\mu) \quad [40]$$

For λ=0, the area is zero. It is convex in the vicinity of zero, and growing everywhere, going toward infinity when λ grows toward infinity. But λ is less than 1 in this context. In general, this function is not necessarily convex, but it becomes convex if σ is sufficiently small. The point I is moved backward with respect to its default position, and the lateral correction when C is sufficiently large continues to follow the prior of roundness. By considering the form of *Fe'*, we expect a less spectacular effect of a decreasing apparent diameter of the Moon as its elevation keeps growing (as the minimum is not far from λ=1 and then μ=1) than with *Fe* above.

The above formulas are valid as well for meridians other than the vertical.

**c) The conformal model**

Let us suppose that we introduce a factor 2 in the formula for f. In the simplest case this yields the following transformation g of the sphere at infinity: Θ=2 arctan($\lambda tan\theta/2$), Φ=φ.

The transformation may look similar to the original f, but it is misleading. The original f sends the hemisphere x=cos(θ) positive into itself, and when extended by continuity to the great circle θ=π/2 (where the tangent is infinite), this extended formula induces an identity transform (i.e. sending points to themselves) on this great circle. It then induces a smooth transformation of the projective plane into itself: the new g is a well-defined smooth map from the whole sphere, x positive and x negative, that fixes the two poles x=1 and x=-1, and sends the equator θ=π/2 to the circle Θ=2 arctan(λ). It is a conformal map for the spherical metric. The division by two of the angles corresponds to the coordinates on the unit sphere viewed from the pole x=-1, and the passage to the tangent corresponds to the stereographic map, which is conformal.

The transformation g is named a *boost* in Special Relativity, because it describes the effect of a boost on the sphere of the past towards the present (or future). Then its intervention in a theory of perception of the sky around us might make sense and may be less surprising than a purely projective map such as f.

In fact, such map g (and others having the same structure) can naturally be defined in the PCM.

The fact that we are looking all around us, although not at the same time, makes sensible a two-fold cover S(V) of the plane at infinity P(V), and the fact that we are able to sometimes sense differences in depths very far, even if we cannot in general attribute a quantitative measure to these differences, renders natural the possibility of spheres associated with S(V) that are very far but remain at a finite distance. On such spheres, as on S(V), the angular metric would be sensed by ordinary vision. Note that, in this context, a change of projective frame can induce a transport of this metric to another one, thus influencing the perception of objects, viewed as surfaces on this sphere and as volumes in general.

Let us take again the default frame in the 4-dimensional space W: $e_1$ in the direction of H, $e_2$ lateral, $e_3$ vertical, and $e_4$ from the point of view O. The 3-dimensional affine space E is the hyperplane containing O and parallel to $e_1$, $e_2$ and $e_3$. Let us also consider a sphere $S_c$ of center O and radius c, and its isomorphic image $P(S_c)$ in the projective space P(W). The subgroup of the group GL(W) of linear bijections of W, constituted by the transformations sending $P(S_c)$ into itself, is the conformal group of the standard Lorentzian metric, which is defined by the quadratic form:

$$Q(x_1, x_2, x_3, x_4) = x_1^2 + x_2^2 + x_3^2 - c^2 x_4^2 \qquad [41]$$

When considering the projective transformations, this yields the group PO(1,3), which is isomorphic to the conformal group of the sphere $S_R$, of which the connected component of the identity is $PSL_2(\mathbf{C})$, when $S_c$ (or $P(S_c)$) is parametrized by a complex number, after stereographic projection.

Let us introduce the *light coordinates* x'=$x_1$+c$x_4$, x"=$x_1$-c$x_4$ (c is not necessarily the speed of light in this context, it denotes any large radius), so that: $x'x'' = x_1^2 - c^2 x_4^2$. The *standard boost* of ratio λ is defined, in homogeneous coordinates, by the transformation y'=x'/λ and y"=λx". It belongs to the connected Lorentz group SO(1,3), and it induces on the sphere $S_c$ the transformation g just described below.

We have:

$y_1$=½(y'+y")=½(($x_1$+c$x_4$)/λ+λ($x_1$-c$x_4$))=½(λ+1/λ)x_1-½(λ-1/λ)c$x_4$   [42]

and,

c$y_4$=½(y'-y")=½(($x_1$+c$x_4$)/λ-λ($x_1$-c$x_4$))=-½(λ-1/λ)$x_1$+½(λ+1/λ)c$x_4$   [43]

but $y_2$=$x_2$, and $y_3$=$x_3$.

Let us define the vector e'=($e_1$+c$e_4$)/2c along the axis of x', and the vector e"=($e_1$-c$e_4$)/2c along the axis of x". The change from $e_1$, $e_4$ to this new basis yields the formulas of the coordinates x', x". These vectors are sent respectively to λe' and e"/λ. Therefore the vector $e_1$=c(e'+e") is sent to $f_1$=½ (λ+1/λ)$e_1$+½(λ-1/λ)c$e_4$, and the vector $e_4$=e'-e" is sent to $f_4$=½(λ-1/λ)$e_1$/c+½(λ+1/λ)$e_4$. The vectors $e_2$ and $e_3$ are fixed. The fifth vector $e_0$=$e_1$+$e_4$ +$e_2$+$e_3$ is sent to:

$f_0$=$f_1$+$f_2$+$f_3$+$f_4$=½((1+1/c)λ+(1-1/c)/λ))($e_1$+$e_4$)+$e_2$+$e_3$   [44]

The inverse of the above transformation is the projective transformation T, which corresponds to the following equations, where now e'$_1$, e'$_2$, e'$_3$, e'$_4$ denotes the image of the original basis:

$$e'_1 = \frac{1}{2}\left(\lambda + \frac{1}{\lambda}\right) e_1 - \frac{1}{2}\left(\lambda - \frac{1}{\lambda}\right) ce_4 \qquad [45]$$
$$e'_4 = -\frac{1}{2c}\left(\lambda - \frac{1}{\lambda}\right) e_1 + \frac{1}{2}\left(\lambda + \frac{1}{\lambda}\right) e_4 \qquad [46]$$
$$e'_2 = e_2 \qquad [47]$$
$$e'_3 = e_3 \qquad [48]$$

We can see that, in P(V), the point O and H are both moved along the line OH. If λ is larger than 1, the segment OH expands, O' being in the back and H' moving backward too. If λ is smaller than 1, the contrary happens, O' and H' both move forward along the segment OH. However, this does not imply that the observer has moved. The change concerns the integration of external variables, as in the preceding section.

Under T or f, the plane at infinity P(V) is not preserved. If λ is larger than 1, infinity moves far away in front, which might correspond to the reported feeling of being closer to the Moon. But if λ is smaller than 1, infinity in front moves closer to the observer, and approaches the sphere of radius c, which might correspond to the opposite feeling that the Moon is farther.

To describe the transform g of the sphere $S_r$ in the 3-dimensional affine Euclidean space E, we have to normalize the equations by imposing $x_4=y_4=1$, which yields the following non-linear formulas:

$$y_1 = c \frac{\left(\lambda+\frac{1}{\lambda}\right)x_1+c\left(\frac{1}{\lambda}-\lambda\right)}{\left(\frac{1}{\lambda}-\lambda\right)x_1+c\left(\lambda+\frac{1}{\lambda}\right)} \qquad [49]$$

$$y_2 = \frac{2cx_2}{\left(\frac{1}{\lambda}-\lambda\right)x_1+c\left(\lambda+\frac{1}{\lambda}\right)} \qquad [50]$$

$$y_3 = \frac{2cx_3}{\left(\frac{1}{\lambda}-\lambda\right)x_1+c\left(\lambda+\frac{1}{\lambda}\right)} \qquad [51]$$

The lateral points $R_c$, $L_c$ as the vertical point $V_c$ on the sphere $S_r$ are moved by T in the direction opposite to H. But the points at infinity, V and L remain fixed. These effects might induce sensations of enlargement of the sky when $\lambda \geq 1$, and the opposite effect when $\lambda \leq 1$.

The analysis of the free energy $Fe(\lambda)$ for this case of finite radius c is not fundamentally different from the case of infinite radius of the celestial sphere as described in the preceding section. Identical arguments lead to the same kind of functional. The areas that we have to consider are again the areas of regions of the form θ smaller than a certain value.
The region of the sphere where θ is larger than π/4, being the reference by default, is sent now by g on the region where $\cos(\Theta)/2$ is larger than $1/\sqrt{(1+\lambda^2\tan^2(\pi/8))}=1/(1+(1+\frac{1}{2}\sqrt{3})\lambda^2)$. Then $\cos(\Theta)$ itself has to be greater than $-1+2/(1+(1+\frac{1}{2}\sqrt{3})\lambda^2)$, and the inverse of the area is now:

$$\rho_c(\lambda)= (\sqrt{2}-1)(1+(1+\sqrt{3}/2)\lambda^2))/2\sqrt{2}(1+\sqrt{3}/2)\lambda^2) \qquad [52]$$

Note that we measure ratios of areas, thus the radius c doesn't enter the formula; the metric could be the angular one as well.
Then the function to minimize when the Moon is between the horizon and π/4, is:

$$Fe(\lambda)= \frac{1}{2\sigma^2}(\lambda-1)^2 +C' \Lambda\rho(\lambda), \qquad [53]$$

as considered for $\lambda \geq 1$.

The complementary function for the Moon high in the sky is:

$$Fe'_c(\lambda)= \frac{1}{2\sigma^2}(\lambda-1)^2 +C'' \Lambda' \left(1+\sqrt{3}/2)\lambda^2\right)/(1+(1+\sqrt{3}/2)\lambda^2)) \qquad [54]$$

when $\lambda \leq 1$.
The group of projective transformations that preserve globally the sphere $P(S_c)$ is the group of isometries of a family of hyperbolic metrics on the ball $P(B_c)$ in P(W) that is contained in $P(S_c)$. Thus it is natural to consider this ball as equipped with one of the Riemannian geometries. An elegant model of this metric is the metric induced by the Lorentz form Q on one of the branches of the hyperboloid $Q(x)=-\rho^2$, where ρ is a strictly positive number. For convenience, we will consider the branch $H_\rho$ with $x_4$ strictly positive too. The rays from 0 in W to the points $H_\rho$ describe the ball $P(B_c)$. A parametrization of this branch is given by:

$$x_1 = \rho \sinh \xi \cos \theta \qquad [55]$$

$x_2 = \rho \sinh \xi \sin \theta \cos \varphi$    [56]
$x_3 = \rho \sinh \xi \sin \theta \sin \varphi$    [57]
$x_4 = \frac{\rho}{c} \cosh \xi$    [58]

In those coordinates ξ,θ,φ, the Riemannian metric takes the following form:

$$ds^2 = \rho^2 (d\xi^2 + \sinh \xi^2 \, (d\theta^2 + \sin \theta^2 \, d\varphi^2)) \qquad [59]$$

The corresponding parametrization of the Euclidean ball $B_c$ in E is obtained by taking the projection to $x_4=1$ and:

$x_1 = c \tanh \xi \cos \theta$    [60]
$x_2 = c \tanh \xi \sin \theta \cos \varphi$    [61]
$x_3 = c \tanh \xi \sin \theta \sin \varphi$    [62]

Then we set r= ctanh(ξ), and get the usual spherical polar coordinates r, θ, φ on the Euclidean space V, identified with E, with the following metric, which is only defined inside the ball $B_c$:

$$ds^2 = \rho^2 (\frac{c^2}{(c^2-r^2)^2} dr^2 + \frac{r^2}{c^2+r^2} (d\theta^2 + \sin \theta^2 \, d\varphi^2)) \qquad [63]$$

When c tends to +∞, this metric, once multiplied by $c^2$, converges to the standard Euclidean metric with scale ρ. The groups of isometries of the hyperbolic metrics, accordingly, tend to the Euclidean group of displacements, similar to a sort of zooming. Thus for large c, we can see the above hyperbolic geometry as a deformation of the ordinary Euclidean geometry.

We cannot exclude that this hyperbolic geometry, which is in exact agreement with the conformal transformation on the large sphere $S_c$ in E, plays a role in perception. In fact, if the conformal change of projective frame is used for perceiving the celestial sphere in some circumstances, this could justify the use of the hyperbolic geometry.

However, this is not necessary in our context to appeal to a hyperbolic geometry, as we do not have to appeal to the Euclidean geometry at finite distance in the first model of f.

**d) Discussion of Heelan's model**

The above family of hyperbolic geometries were introduced in the domain of visual perception by P.A. Heelan *(17)*. The name he gave to this theory was the "hermeneutic Luneburg model", because Rudolf Luneburg (see *17*) developed the idea that available primary cues and priors, together with sensorimotor constraints, could be expressed by a change in the geometry of perception. Thus it should be evident that the approach of Luneburg and Heelan is an ancestor of our approach.

However, there are many fundamental differences. First, let us consider differences of detail in the setting. Heelan finds the above $ds^2$ in bipolar coordinates, which represent a deformation of spherical polar coordinates, in order to take into account the disparity in vision induced by the two eyes (*17*, pp. 286-287; we take the opportunity here to note a typographic error in the formula of the metric page 287, where the factor $\sinh(\xi)^2$ seems to appear at a wrong place). We note that Heelan was in general more interested in near vison in a room than in far vision. Also, in the variables, Heelan noted ρ=κ (which changes nothing), but he also noted: ξ= σγ+στ; which is more important. The variable coordinate is γ, and σ,τ are constants, the first indicating the Euclidean distance from the observer to the region of the physical space where a good agreement between the hyperbolic metric and Euclidean metric can be expected. (Note that Heelan says that τ is for the curvature, but this cannot be what he had in mind, because no change of variables has an effect on the curvature, since it is intrinsic and controlled here by the inverse of ρ=κ.)

But there are more important divergences between our point of view and Heelan's. In our model, the metric results from two choices: a frame (projective in nature) and a kind of measurement (angles in the present case, but this could be another metric in another case). In Heelan, a change of frame is not considered and the metric is chosen *a priori* in a fixed family. Moreover, for the choice of parameters, Heelan compared the perceptual metric with the physical metric somewhere in space, which in principle results from a frame and does not precede it. In some sense, real physical space has a stronger status in the approach of Heelan. This appears in his use of the term "true point" for the region where $ds^2$ must be Euclidean.

Of course, parameters in the metric and parameters in the frames are connected. Moreover, we cannot neglect that the free energy, which controls the change of frame in our model, also uses geometric parameters as priors (such as ordinary distances inside Kullback-Leibler divergence, or areas for entropy). Thus these differences could be somewhat superficial. However, in our model, we can make a clear connection between geometry and information content, which is more obscure in Heelan's.

As discussed in the main text, with respect to the particular perceptual situation of the Moon Illusion, our model is compatible with conformal constraints, which can maintain the round shape of the Moon, but it also offers possible non-conformal solutions. Thus the PCM is compatible with possible elliptical effects in the perception of the shape of the Moon. Any departure from conformality would not be accounted for by Heelan's model (see main text).

Last but not least, our model, while being capable of integrating the essential properties captured by Heelan's model, has much broader explanatory and predictive power, as it encompasses and unifies perception, imagination, and motor programming. This is in parts due to the characteristics of projective geometry, which is not attached to a particular family of metrics but rather represents an extended notion of variations of points of view. In particular, the convenient families of metrics which could be considered at finite distances, in other contexts, are not restricted to hyperbolic ones; they can be spherical or Euclidean. A more general model with broader explanatory and predictive power should be selected over a model that is equivalent to the broader model in a very specific context but that cannot account for a wealth of essential phenomena that are linked together experientially and unified in the more general model.

**e) Further remarks on the default frames**

**Choice by default of the fifth point I**

It could be that some persons prefer to place the point I to their right and others to their left, just as there are right-handed persons and left-handed persons. However, it could also happen that most persons use two points $I_R$, $I_L$ symmetric with respect to the sagittal plane. This pair would imply more than one frame, but the theory can easily be extended to this kind of enriched frame. The default 45° situation of $I_R$, $I_L$ or I with respect to the sagittal plane is compatible with the orientations at 45° of the planes of semi-circular canals in the labyrinths of vertebrates and the corresponding orientations of the eye muscles. These two possibilities, one point or a symmetric pair, could be tested experimentally by manipulating left and right environmental cues. This is also something that could be empirically tested in future studies.

**Analogy with color adaptation**

The geometry governing color perception is fairly complex. However, as a first approximation, 3D affine geometry explains well the change in perception induced by a change in illumination. Here, the natural or "default" frame is based on the distribution of wave-lengths in solar light, related to the center and shapes around it, the three axes corresponding to luminance S+L+M, to green/red opposition, L-M, and to blue/yellow opposition, S-L-M. When the light distribution is modified, the origin, shape, and coordinates must change accordingly in order to maintain sufficient information flow.